\documentclass[fleqn,superscriptaddress,showpacs,nofootinbib]{revtex4}

\usepackage{amssymb,amsmath,epsfig}

\DeclareMathOperator{\tr}{Tr}

\def\slash#1{\setbox0=\hbox{$#1$}               % set a box for #1
        \dimen0=\wd0                            % and get its size
        \setbox1=\hbox{/} \dimen1=\wd1          % get size of /
        \ifdim\dimen0>\dimen1                   % #1 is bigger
        \rlap{\hbox to \dimen0{\hfil/\hfil}}    % so center / in box
        #1                                      % and print #1
        \else              
        \rlap{\hbox to \dimen1{\hfil$#1$\hfil}} % so center #1
        /                                       % and print /
        \fi}                                    %

\begin{document}

\title{Dijet imbalance in hadronic collisions}

\author{Dani\"el Boer}
\email{D.Boer@kvi.nl}
\affiliation{
Department of Physics and Astronomy, Vrije Universiteit Amsterdam\\
NL-1081 HV Amsterdam, The Netherlands}
\affiliation{Theory Group, KVI, University of Groningen\\ 
Zernikelaan 25, NL-9747 AA Groningen, The Netherlands}

\author{Piet J. Mulders}
\email{mulders@few.vu.nl}
\affiliation{
Department of Physics and Astronomy, Vrije Universiteit Amsterdam\\
NL-1081 HV Amsterdam, The Netherlands}

\author{Cristian Pisano}
\email{cristian.pisano@ca.infn.it}
\affiliation{
Department of Physics and Astronomy, Vrije Universiteit Amsterdam\\
NL-1081 HV Amsterdam, The Netherlands}

\affiliation{Dipartimento di Fisica, Universit\`a di Cagliari,\\ 
and INFN, Sezione di
Cagliari, I-09042 Monserrato (CA), Italy}

%%%%%%%%%%%%%%%%%%%%%%%%%%%%%%%%%%%%%%%%%%%%%%%%%%%%%%%%%%%%%%%%%%%%%%%%%
%%%%%%%%%%%%%%%%%%%%%%%%%%%%%%%%%%%%%%%%%%%%%%%%%%%%%%%%%%%%%%%%%%%%%%%%%
\begin{abstract}
  The imbalance of dijets produced in hadronic collisions has been
  used to extract the average transverse momentum of partons inside
  the hadrons. In this paper we discuss new contributions to the dijet
  imbalance that could complicate or even hamper this extraction.
  They are due to polarization of initial state partons inside
  unpolarized hadrons that can arise in the presence of nonzero parton
  transverse momentum. Transversely polarized quarks and linearly
  polarized gluons produce specific azimuthal dependences of the two
  jets that in principle are not suppressed. Their effects cannot be
  isolated just by looking at the angular deviation from the
  back-to-back situation, rather they enter jet broadening
  observables. In this way they directly affect the extraction of the
  average transverse momentum of unpolarized partons that is thought
  to be extracted. We discuss appropriately weighted cross sections to
  isolate the additional contributions.
\end{abstract}

\date{\today}

\maketitle
%%%%%%%%%%%%%%%%%%%%%%%%%%%%%%%%%%%%%%%%%%%%%%%%%%%%%%%%%%%%%%%%%%%%%%%%%%
%%%%%%%%%%%%%%%%%%%%%%%%%%%%%%%%%%%%%%%%%%%%%%%%%%%%%%%%%%%%%%%%%%%%%%%%%%
%%%%%%%%%%%%%%%%%%%%%%%%%%%%%%%%%%%%%%%%%%%%%%%%%%%%%%%%%%%%%%%%%%%%%%%%%%
\section{Introduction}
Event shape observables have been widely studied for various reasons.
In $e^+ e^-$-annihilation, observables such as the thrust and
jet broadening have been studied primarily to extract $\alpha_s(M_z)$, cf.\
for instance
Refs.\ 
\cite{Rakow:1981qn,Catani:1992jc,Dokshitzer:1998kz,Jones:2003yv,GehrmannDeRidder:2007hr,Weinzierl:2009ms}
for theoretical studies and Refs.\ 
\cite{Berger:1983yp,Naroska:1986si,Acton:1993zh,Abe:1994mf,MovillaFernandez:1997fr,Pfeifenschneider:1999rz,Abdallah:2004xe,Achard:2004sv,Abbiendi:2005gb} for experimental studies. In
the center-of-mass system (cms) 
of the $e^+ e^-$ collisions at lowest order in $\alpha_s$, the
produced quark-antiquark pair is exactly back-to-back leading for
two-jet events to a thrust $T$
equal to unity. Gluon radiation, i.e.\ order $\alpha_s$ corrections,
gives rise to nonzero $1-T$ and also to nonzero jet broadening. In the
perturbative regime these observables can be used to extract
$\alpha_s$, which has been done recently at next-to-next-to-leading order 
\cite{Dissertori:2007xa,Bethke:2008hf,Dissertori:2009qa}.
The results compare very well with those obtained by other means
of extraction. In the nonperturbative regime, hadronization will also lead
to nonzero event or jet shapes. This is
characterized by a mean transverse momentum $\langle k_\perp \rangle$, leading
in general to a contribution suppressed by a power of the large scale,
the cms energy $Q$. For example, in the nonperturbative regime
$1-T\propto \langle k_\perp \rangle/Q$. It has been suggested
that this contains universal information on $\alpha_s$ in the
infrared regime. We refer to Ref.\ \cite{Dokshitzer:1998nz} for a review
on this topic.

Event and jet shapes have also been studied in hadronic collisions.
Compared to $e^+ e^-$-annihilation here the additional complication
of initial parton transverse momenta arises. Another difference is that instead
of the thrust axis, it is common to use the transverse thrust axis $n_t$,
which is the axis in the transverse plane having maximum transverse energy
flow. The corresponding transverse thrust is defined as
\cite{Ellis:1986ig}: 
\begin{equation}
T_t= \max \sum_{i=1}^{n} \frac{|\boldsymbol{p}_T^i \cdot \boldsymbol{n}_t|}{E_T},
\end{equation}
where $\boldsymbol{p}_T^i$ is the transverse momentum of the outgoing hadron $i$,
$E_T=\sum_i |\boldsymbol{p}_T^i|$ is the total transverse energy 
(neglecting masses)
and the transverse thrust axis is the transverse unit vector $\boldsymbol{n}_t$ that
maximizes $T_t$. Here we use the notation of Ref.\ \cite{Ellis:1986ig}, where
also the jet broadening variable $Q_t$ is defined as:
\begin{equation}
Q_t= \sum_{i=1}^{n} |\boldsymbol{p}_T^i \times \boldsymbol{n}_t|.
\label{eq:Qt}
\end{equation}
An experimental investigation of the average $Q_t$ as a function of $E_T$
in $p \bar{p}$ collisions has been reported in Ref.\
\cite{Abe:1991dma}. Higher order perturbative corrections to
the transverse thrust and jet broadening are
discussed in e.g.\ Refs.\ \cite{Nagy:2003tz,Banfi:2004nk}.

Assuming collinear factorization and ignoring broadening from hadronization,
$Q_t$ will be zero for $2 \to 2$ partonic subprocesses and only sensitive to
$2 \to 3$ processes, like for $e^+ e^-$-annihilation except that
there are more subprocesses to consider in hadronic collisions. 
Extraction of $\alpha_s$ in hadronic collisions
\cite{Ellis:1992qq,Affolder:2001hn} is however
complicated due to the presence of parton transverse momenta and
the transverse momentum distribution of hadrons inside the jet. The former
effect one
can minimize by considering events with at least three pronounced jets, which
means considering only large values of $Q_t$, whereas
the latter effect could be minimized
by considering $Q_t$ for jets, instead of hadrons.
In fact, the quantity $Q_t$ for two-jet events, where $i$ now denotes the 
$i$-th jet and $n=2$, has been used to study and 
extract the average parton transverse momentum. This has been done for
instance in Refs.\ 
\cite{Angelis:1978uv,Clark:1979vc,Baier:1979tp,Corcoran:1979kd,Angelis:1980bs,Begel:1999rc,Levai:2005fa,Adler:2006sc,Morsch:2006pf,Fai:2006pd}.
As can be seen from those results, the average parton transverse momentum
extracted from the data increases with energy ($\sqrt{s}$) and is in general
much too large to be attributable to 
``intrinsic'' transverse momentum. This is 
a consequence of soft parton radiation, similar to what happens for the
transverse momentum distribution of vector boson production in hadronic
processes
\cite{Altarelli:1977kt,Parisi:1979se,Collins:1984kg,Altarelli:1984xd} 
(see also the instructive discussion in Ref.\ \cite{Collins:1985kw}). 
Resummation of soft radiation 
effectively broadens the transverse momentum
dependence of the parton distributions, increasingly so with 
increasing center of mass energy. 

In this paper we point out
that besides initial parton transverse momentum and soft parton
radiation, there are additional contributions to $Q_t$, even for
the simplest two-jet case. These are 
contributions due to the transverse polarization of quarks and the
linear polarization of gluons inside the initial {\em unpolarized\/} hadrons.
These contributions can arise for
nonzero initial parton transverse momentum. We will show how these
effects contribute to $Q_t$ and discuss that besides complicating the
extraction of the average parton transverse momentum from $Q_t$,
they may even hamper that extraction altogether depending on their magnitude.

In Ref.\ \cite{Ellis:1986ig} collinear factorization was
assumed, making the observable $\langle Q_t \rangle$ only sensitive to $2 \to
3$ subprocesses.
In reality collinear factorization is not always applicable, due to the
partonic transverse momentum effects.
In a simple picture of a Gaussian distribution
of intrinsic parton momentum $p_t$, the average value $\langle p_t \rangle$
can be extracted from $\langle Q_t \rangle$, but in fact, no factorization
theorem has been established for two-jet or two-hadron production in
$p p$ or $p \bar{p}$ collisions for observables that are sensitive to
parton transverse momenta. To make matters worse, in the framework of
transverse momentum dependent parton distribution functions, nowadays commonly
referred to as TMDs, it even seems that factorization cannot be
established for this particular type of process when taking into
  account nontrivial effects of gauge links
\cite{Bomhof:2006dp,Pijlman:2006tq,Collins:2007nk,Collins:2007jp,MuldersRogers09}.
This would cast doubt on any conclusion
drawn from $\langle Q_t \rangle$ in hadronic collisions,
except for large $Q_t$ where collinear factorization {\it can} be applied.
But even if
factorization will work out in some as yet unknown way, the additional
contributions from spin dependent TMDs may complicate matters considerably. 
Schematically this can be seen as follows.

Consider the process $h_1 h_2 \to j_1 j_2 X$, where $j_i$ stands for
produced jet $i$.
In the plane transverse to the collision axis, $\delta
\phi$ denotes the deviation of the
(azimuthal) angle between the two jets from $\pi$,
i.e.\ $\delta \phi=\phi_{j_1}-\phi_{j_2}-\pi$. It is sometimes referred to as
the dijet imbalance. Let us consider only $2 \to 2$ subprocesses.
In collinear factorization the $\delta \phi$ dependence of the cross section
will then only receive a contribution at $\delta \phi=0$. 
Allowing for parton transverse momentum in the initial hadrons leads to a
smearing of the $\delta \phi$ distribution. 
For the idealized case of equal jet transverse momenta (both
equal to $E_T/2$) the differential cross section takes the form: 
\begin{equation}
\frac{d\sigma}{dE_T d\delta\phi}= A(Q_t^2) + B(Q_t^2) Q_t^2 +
C(Q_t^2) Q_t^4\, ,
\label{schematicxs}
\end{equation}
where $Q_t=E_T |\sin(\delta \phi/2)|$ is equal to the absolute value of the
transverse momentum of the two-jet system. $A,B,C$ are functions of $Q_t^2$,
which do not need to vanish at $Q_t^2=0$. The terms $B$ and $C$
appear from spin effects
inside the initial hadrons $h_i$, for which expressions will be
presented in this paper. In general these spin-dependent
  contributions are not suppressed by powers of $1/E_T$, also not when
  arising from polarized gluons as claimed in Ref.\ \cite{Lu:2008qu}.
A result for $B$ has recently been obtained in \cite{Lu:2008qu}
following a calculation similar to the one
for $p \bar{p} \to \gamma j X$ presented in \cite{Boer:2007nd}.
This contribution arises from the quark
TMD $h_1^{\perp\, q}$ \cite{Boer:1997nt}, which represents the distribution of
transversely polarized quarks inside an unpolarized hadron.
The new result in this
paper is the contribution from $h_1^{\perp\,g}$ \cite{Mulders:2000sh}, the
distribution of linearly polarized gluons inside an unpolarized hadron, which 
gives rise to $C$.
Upon ignoring these spin effects, only the term $A$ remains and the average
$Q_t$ value in that case will indeed be directly related to the average
transverse momentum that is thought to be extracted in Refs.\ 
\cite{Angelis:1978uv,Clark:1979vc,Baier:1979tp,Corcoran:1979kd,Angelis:1980bs,Begel:1999rc,Levai:2005fa,Adler:2006sc,Morsch:2006pf,Fai:2006pd}. Our results in principle cast doubt on whether the actual 
value of $\langle k_t^2 \rangle$ has been extracted in those cases. In
practice, it all depends on the magnitude of $B$ and $C$. We will present a
simple Gaussian model to illustrate the generic shape of the 
modification of the dijet imbalance distribution by $B$
and $C$ terms. 
 
The paper is organized as follows. First we will present the calculation and
expressions for the cross section in Eq.\ (\ref{schematicxs}), 
assuming factorization in terms of transverse momentum dependent
correlators and ignoring the possible effects from gauge links. We will
actually discuss the more general case in which the two jet transverse 
momenta are
not equal, but differ by a small amount w.r.t.\ $E_T$. 
In that case the angular dependence is
more involved than given in Eq.\ (\ref{schematicxs}), even upon expansion 
in the small transverse momentum difference of the two jets with respect to 
their sum. We will first express the cross section in terms of
the individual jet momenta through their sum and difference 
(section~\ref{section2}, in particular Eq.\ (\ref{eq:cso})) 
and subsequently in terms of the sum and
difference of the lengths of the jet momenta in order to 
arrive at the dijet imbalance distribution expressed in more standard
variables (section~\ref{sec:imbalance}, in particular Eq.\ (\ref{xsdeltaphi})).
In section~\ref{sec:weightedxs} we discuss angular-projected asymmetries, 
such as $\langle\cos\delta \phi \rangle$ and the ones that can be
used to extract $B$ and $C$. After that we consider
the consequences of nonzero $h_1^\perp$ functions 
for the jet broadening quantity $Q_t$, in particular for the averages
$\langle Q_t \rangle$ and $\langle Q_t^2\rangle$. 
Finally (section~\ref{section6}) we briefly address the open issues
of {factorization (breaking) and color flow dependence 
upon inclusion of gauge links. 
We end with conclusions and two appendices, one on relations among
various variables in the transverse plane and one on photon-jet
production that completes the treatment given in Ref.\ \cite{Boer:2007nd}.
  
%%%%%%%%%%%%%%%%%%%%%%%%%%%%%%%%%%%%%%%%%%%%%%%%%%%%%%%%%%%%%%%%%%%%%%%%%%
%%%%%%%%%%%%%%%%%%%%%%%%%%%%%%%%%%%%%%%%%%%%%%%%%%%%%%%%%%%%%%%%%%%%%%%%%%
\section{\label{section2}
Theoretical framework: calculation of the cross section}
%%%%%%%%%%%%%%%%%%%%%%%%%%%%%%%%%%%%%%%%%%%%%%%%%%%%%%%%%%%%%%%%%%%%%%%%%%
%%%%%%%%%%%%%%%%%%%%%%%%%%%%%%%%%%%%%%%%%%%%%%%%%%%%%%%%%%%%%%%%%%%%%%%%%%

We consider the process 
\begin{equation}
h_1(P_1){+}h_2(P_2)\, {\rightarrow}\,{\rm jet}(K_1){+}{\rm jet}(K_2){+}X \, ,
\end{equation}
where the four-momenta of the particles are given within brackets, and
the jet-jet pair in the final state is almost back-to-back in the plane 
perpendicular to the direction of the incoming hadrons. 
Along the lines of Ref.\ \cite{Boer:2007nd}, we will instead of collinear 
factorization
consider a generalized factorization scheme taking into account 
partonic transverse momenta.
We make a lightcone decomposition of the two incoming hadronic momenta in terms of the light-like Sudakov vectors $n_+$ and $n_-$, satisfying $n_+^2\,{=}\,n_-^2\,{=}\,0$ and $n_+{\cdot}n_-\,{=}\,1$:
\begin{equation}
P_1^\mu
=P_1^+n_+^\mu+\frac{M_1^2}{2P_1^+}n_-^\mu\ ,\qquad\text{and}\qquad
P_2^\mu
=\frac{M_2^2}{2P_2^-}n_+^\mu+P_2^-n_-^\mu\ ~.
\end{equation}
The partonic momenta ($p_1$, $p_2$) can be expressed  in terms of the  
lightcone momentum fractions ($x_1$, $x_2$) and the 
intrinsic transverse momenta ($ p_{1 T}$, $ p_{2 T}$), as follows
\begin{equation}
p_1^\mu
=x_1^{\phantom{+}}\!P_1^+n_+^\mu
+\frac{p_1^2{+}\boldsymbol p_{1 T}^2}{2x_1^{\phantom{+}}\!P_1^+}n_-^\mu
+p_{1 T}^\mu\ ,
\qquad\text{and}\qquad
p_2^\mu
=\frac{p_2^2{+}\boldsymbol p_{2 T}^2}{2x_2^{\phantom{-}}\!P_2^-}n_+^\mu
+x_2^{\phantom{-}}\!P_2^-n_-^\mu+p_{2 T}^\mu\ .
\label{PartonDecompositions}
\end{equation}
In general $n_+$ and $n_-$ will define the lightcone components of every
vector $a$ as $a^\pm \equiv a \cdot n_\mp$, while 
perpendicular vectors $a_\perp$  will always refer to the components of  
 $a$ orthogonal to both incoming hadronic momenta, $P_1$ and $P_2$. 
Therefore in Eq.\ (\ref{PartonDecompositions}), if we neglect hadron masses, 
$p_{1 T}^{\mu}= p_{1 \perp}^{\mu}$
and $p_{2 T}^{\mu}= p_{2 \perp}^{\mu}$.
We denote with $s$ the total energy squared in the hadronic 
cms frame, $s =(P_1+P_2)^2 = E^2_{\rm cms} $, and 
 with $\eta_i$  the pseudo-rapidities 
of the outgoing partons,
\emph{i.e.}\ $\eta_i\,{=}\,{-}\ln\big(\tan(\frac{1}{2}\theta_i)\big)$,  
$\theta_i$ being the polar angles of the outgoing partons in the same frame. 
Finally, we introduce the  partonic Mandelstam variables  
\begin{equation}
\hat s = (p_1 + p_2)^2, \qquad \hat t = (p_1-K_1)^2 ,\qquad 
\hat u = (p_1-K_2)^2, 
\end{equation}
which satisfy the relations
\begin{equation}
\label{Yexpression}
 -\frac{\hat t}{\hat s} 
\equiv y = \frac{1}{e^{\eta_1 -\eta_2}\,{+}\,1}~ , \qquad {\rm and} \qquad   -\frac{\hat u}{\hat s} = 1-y~.
\end{equation}

Following Refs.\ \cite{Boer:2007nd} and \cite{Bacchetta:2007sz}  
we assume that at sufficiently high energies the hadronic cross section 
factorizes in a soft parton correlator for each observed hadron and a hard 
part:
\begin{eqnarray}
d\sigma^{h_1 h_2 \rightarrow {\rm jet} \,{\rm jet} \, X}
& = &\frac{1}{2 s}\, \frac{d^3 K_1}{(2\pi)^3\,2 E_1}
\frac{d^3K_2}{(2\pi)^3\,2E_2}
{\int}d x_1\, d^2\boldsymbol p_{1 T}\,d x_2\, d^2\boldsymbol p_{2 T}\, (2\pi)^4
\delta^4(p_{1}{+}p_{2}{-}K_{1}{-}K_{2})
 \nonumber \\
&&\qquad \qquad\qquad \qquad\qquad\times \sum_{a{,} b{,} c{,} d}\ 
\Phi_a(x_1{,}\boldsymbol p_{1 T})\otimes\Phi_b(x_2{,}\boldsymbol p_{2 T})
\otimes\,|H_{a b  \rightarrow c d }(p_1, p_2, K_{1}, K_2)|^2\ .
\label{CrossSec}
\end{eqnarray}
This form assumes the simplest possible factorization omitting
any gauge link dependence in the correlators, which can modify or
even break the factorization (see section \ref{section6}
for a discussion of these open issues).

In Eq.\ (\ref{CrossSec}) the sum runs over all the incoming and
outgoing partons taking part in the reaction.
The convolutions $\otimes$ indicate the appropriate traces over Dirac 
indices and $|H|^2$ is the hard partonic squared amplitude. The parton
 correlators are  defined on the lightfront LF ($\xi{\cdot}n\,{\equiv}\,0$, with $n\equiv n_-$ for parton 1 and $n\equiv n_+$ for parton 2);  they describe 
the hadron $\rightarrow$ parton transitions   and can be 
parameterized in terms 
of transverse momentum dependent (TMD) distribution functions.
In particular, the quark content of an 
unpolarized hadron is at leading twist (omitting gauge links) described 
by the correlator \cite{Boer:1997nt}
\begin{eqnarray}
\label{QuarkCorr}
\Phi_q(x{,} \boldsymbol p_T)
%%%%
=  {\int}\frac{d(\xi{\cdot}P)\,d^2\xi_T}{(2\pi)^3}\ e^{ip\cdot\xi}\,
\langle P |\,\overline\psi(0)\,
\psi(\xi)\,|P\rangle\,\big\rfloor_{\text{LF}}%\nonumber \\
%%%%
=  \frac{1}{2}\,
\bigg \{\,f_1^q(x{,}\boldsymbol{p}_T^2)\;\slash P
+i h_1^{\perp\,q}(x{,}\boldsymbol{p}_T^2)\;\frac{[\slash p_T , 
\slash P]}{2 M}\bigg \}\, , 
\end{eqnarray}
where  $f_1^q(x, \boldsymbol{p}_T^2)$ is the 
unpolarized quark 
distribution, which integrated over $\boldsymbol{p}_T$ gives the familiar 
lightcone momentum distribution $f_1^q(x)$.  
The time-reversal (T) odd function  
$h_1^{\perp q}(x, \boldsymbol{p}_T^2)$ is interpreted as the quark 
transverse spin distribution in an unpolarized hadron
\cite{Boer:1997nt}. Analogously, for an antiquark,
\begin{eqnarray}
\label{AquarkCorr}
\bar\Phi_q(x{,} \boldsymbol p_T)
%%%%
=  -{\int}\frac{d(\xi{\cdot}P)\,d^2\xi_T}{(2\pi)^3}\ e^{-ip\cdot\xi}\,
\langle P |\,\overline\psi(0)\,
\psi(\xi)\,|P\rangle\,\big\rfloor_{\text{LF}}%\nonumber \\
%%%%
=  \frac{1}{2}\,
\bigg \{\,f_1^{\bar q}(x{,}\boldsymbol{p}_T^2)\;\slash P
+i h_1^{\perp\,\bar q}(x{,}\boldsymbol{p}_T^2)\;\frac{[\slash p_T , 
\slash P]}{2 M}\bigg \}\, . 
\end{eqnarray}
The gluon correlator (omitting gauge links) is given by \cite{Mulders:2000sh}
\begin{eqnarray}
\label{GluonCorr}
\Phi_g^{\mu\nu}(x{,}\boldsymbol p_T )
& =&  \frac{n_\rho\,n_\sigma}{(p{\cdot}n)^2}
{\int}\frac{d(\xi{\cdot}P)\,d^2\xi_T}{(2\pi)^3}\ e^{ip\cdot\xi}\,
\langle P|\,\tr\big[\,F^{\mu\rho}(0)\,
F^{\nu\sigma}(\xi)\,\big]
\,|P \rangle\,\big\rfloor_{\text{LF}} \nonumber \\
%%%%
&=&\frac{1}{2x}\,\bigg \{-g_T^{\mu\nu}\,f_1^g(x{,}\boldsymbol{p}_T^2)
+\bigg(\frac{p_T^\mu p_T^\nu}{M^2}\,
{+}\,g_T^{\mu\nu}\frac{\boldsymbol p_T^2}{2M^2}\bigg)\;h_1^{\perp\,g}(x{,}\boldsymbol{p}_T^2) \bigg \}\, ,
\end{eqnarray}
with $g^{\mu\nu}_{T}$ being a transverse tensor defined  as
\begin{equation}
g^{\mu\nu}_{T} = g^{\mu\nu} - n_+^{\mu}n_-^{\nu}-n_-^{\mu}n_+^{\nu}\, .
\end{equation}
The function $f_1^g(x{,}\boldsymbol{p}_T^2)$ represents the 
unpolarized gluon distribution, while the T-even function 
$h_1^{\perp\,g}(x{,}\boldsymbol{p}_T^2)$ is the distribution of
linearly polarized gluons in an unpolarized hadron.

In order to derive an expression for the cross section in terms of parton 
distributions, we insert the parametrizations in Eqs.\ (\ref{QuarkCorr}), 
(\ref{AquarkCorr}) and  (\ref{GluonCorr}) of the TMD 
correlators into Eq.\ (\ref{CrossSec}). Furthermore, utilizing the 
decompositions of the parton momenta in Eq.\ \eqref{PartonDecompositions},  
the $\delta$-function in Eq.\ (\ref{CrossSec}) can be rewritten as
\begin{eqnarray}
\label{DeltaFunc}
\delta^4(p_1{+}p_2{-}K_1{-}K_2)
& = &
\frac{2}{s}\,
\delta\bigg(\,x_1{-}\frac{1}{\sqrt s}
(\,|\boldsymbol K_{1 \perp}|\,e^{\eta_1}\,
{+}|\boldsymbol K_{2 \perp}|\,e^{\eta_2}\,)\,\bigg)\,
\delta\bigg(\,x_2{-}\frac{1}{\sqrt s}
(\,|\boldsymbol K_{1 \perp}|e^{-\eta_1}\,
{+}|\boldsymbol K_{2 \perp}|\,e^{-\eta_2}\,)\,\bigg) \nonumber \\
%%%%
&&\mspace{200mu}
\times\delta^2(\boldsymbol p_{1 T}{+}\boldsymbol p_{2 T}{-}\boldsymbol K_{1 \perp}{-}\boldsymbol K_{2 \perp})\ ,
\end{eqnarray}
with corrections of order $\mathcal O(1/s)$. After integration over $x_1$ and $x_2$, from the first two $\delta$-functions 
on the r.h.s.\ of Eq.\ (\ref{DeltaFunc}), one obtains 
\begin{equation} 
x_1 =\frac{1}{\sqrt s}\bigg(\,|\boldsymbol K_{1 \perp}|\,e^{\eta_1}\,
{+}|\boldsymbol K_{2 \perp}|\,e^{\eta_2}\,\bigg ), \,\quad
x_2{=}\frac{1}{\sqrt s}
\bigg (\,|\boldsymbol K_{1 \perp}|e^{-\eta_1}\,
{+}|\boldsymbol K_{2 \perp}|\,e^{-\eta_2}\, \bigg )\, , 
\end{equation}
which relates the partonic momentum fractions $x_1$, $x_2$ to the rapidities and the transverse momenta of the 
jets. These basic tree-level relations will be used in our treatment. We
will not consider several other effects that need to be accounted for
in practice such as
the actually used jet definition and higher order corrections that affect
the above relations and cause additional smearing.

The hadronic cross section can be written in the form
\begin{equation}
\frac{d\sigma^{h_1 h_2 \rightarrow {\rm jet} \,{\rm jet} \, X}}{d\eta_1 d\eta_2 d^2\boldsymbol{K}_{1\perp} d^2\boldsymbol{K}_{2\perp}} =
\frac{\alpha_s^2}{s \boldsymbol{K}_\perp^2}
\bigg[ A(\boldsymbol{q}_T^2) + B(\boldsymbol{q}_T^2) \boldsymbol q_T^2 \cos 2 (\phi_T-\phi_\perp) + 
C(\boldsymbol{q}_T^2) \boldsymbol{q}_T^4 \cos 4 (\phi_T-\phi_\perp)\bigg ]
\label{eq:cso}
\end{equation}
where $\boldsymbol{q}_T 
\equiv \boldsymbol{K}_{1 \perp} + \boldsymbol{K}_{2 \perp}$ 
and $\boldsymbol{K}_\perp \equiv (\boldsymbol{K}_{1\perp}
- \boldsymbol{K}_{2\perp})/2$. 
The sum momentum $\boldsymbol{q}_T$
is useful as an in principle accessible experimental 
observable momentum which in our calculations via the delta function in
Eq.\ (\ref{DeltaFunc}) is related to intrinsic transverse momenta,
$\boldsymbol{q}_T =  \boldsymbol{p}_{1 T} + \boldsymbol{p}_{2 T}$. 
We denote with $\phi_T$ and $\phi_\perp$
the azimuthal angles of $\boldsymbol{q}_T$ and $\boldsymbol{K}_\perp$,
respectively. 
Besides $\boldsymbol q_T^2$, the terms $A$, $B$ and $C$ depend 
on other kinematic variables often not explicitly indicated, 
namely $y$, $x_1$, $x_2$, and  
contain convolutions of the various parton distributions. These
are discussed separately in the following three subsections, 
where explicit expressions for them can be found, 
calculated at leading order (LO) in perturbative QCD. 
In deriving these expressions we will often employ the approximation 
$|\boldsymbol{q}_T| \ll |\boldsymbol{K}_{1 \perp}| \approx
|\boldsymbol{K}_{2 \perp}| \approx |\boldsymbol{K}_\perp|$  
which is applicable in the situation 
in which the two jets are almost back-to-back in the transverse
plane. However, in deriving Eq.\ (\ref{eq:cso}) we must be
particularly careful with the angular dependence, because 
approximations in the angular dependence that boil down to
approximating $\phi_1 \approx \phi_2 + \pi$ (such as in Eq.\ (21) 
of Ref.\ \cite{Boer:2007nd}) will of course not give the
proper dependence of the dijet imbalance angle 
$\delta\phi = \phi_1 - \phi_2 + \pi$.  
In Eq.\ (\ref{eq:cso}) the combination $\phi_T-\phi_\perp$ appears, which
will allow to isolate the terms $B$ and $C$ by
$\boldsymbol{q}_T^2$-weighted integration over
$\boldsymbol{q}_T$ (cf.\ section \ref{sec:weightedxs}). 
However, in order to arrive at the $\delta \phi$
distribution discussed in the introduction, 
it is more convenient to express the cross section in terms
of the combination $\phi_T-\phi_j$, where $\phi_j$ is the average
jet direction angle, i.e.\ $\phi_j = (\phi_1+\phi_2-\pi)/2$ with $\phi_1$ 
and $\phi_2$ the azimuthal angles of the two outgoing jets 
in the transverse plane. In the present case where
  $|\boldsymbol{K}_\perp| \gg (|\boldsymbol{K}_{1 \perp}| - 
|\boldsymbol{K}_{2 \perp}|$), it holds that $\phi_\perp \approx \phi_j$ 
allowing the two angles to be identified to good approximation for all
values of $\delta \phi$ (cf.\ Eq.\ (\ref{four})).
In the limiting case when $|\boldsymbol{K}_{1 \perp}| =
|\boldsymbol{K}_{2 \perp}|$, the angles $\phi_\perp$ and $\phi_j$ 
exactly coincide
and the $T$ and $\perp$ directions are orthogonal, so we have 
exactly $\cos 2 (\phi_T-\phi_\perp)= -1$ (note that this will lead to 
Eq.\ (\ref{schematicxs}) with a minus sign in front of $B$, but that
is of course only a matter of definition) and 
$\cos 4 (\phi_T-\phi_\perp)=1$. This implies that all angular dependence
then resides in $\boldsymbol{q}_T^2$, which is in that case solely
depends on the off-collinearity of the jets through the dijet imbalance angle
$\delta\phi$ (discussed in section \ref{sec:imbalance}).
 
%%%%%%%%%%%%%%%%%%%%%%%%%%%%%%%%%%%%%%%%%%%%%%%%%%%%%%%%%%%%%%%%%%%%%%%%%%%%%%%
\subsection{Angular independent part of the cross section}
%%%%%%%%%%%%%%%%%%%%%%%%%%%%%%%%%%%%%%%%%%%%%%%%%%%%%%%%%%%%%%%%%%%%%%%%%%%%%%%
The term $A$ in Eq.\ (\ref{eq:cso}) is the angular independent part of the cross 
section and is given by  the sum 
of several contributions ${\cal A}^{a b\to c d}$ coming from the 
partonic subprocesses $a b\to c d$ underlying the reaction $h_1\, h_2\to {\rm jet}\, {\rm  jet }\, X$:
\begin{equation}
A(y, x_1, x_2, \boldsymbol{q}_{T}^2) = \sum_{a, b, c, d} {\cal A}^{a b\to c d}(y, x_1, x_2, \boldsymbol{q}_{T}^2)\, , 
\label{eq:A}
\end{equation}
with $a$,..., $d = q$, $q'$, $\bar q$, $\bar q'$, $g$. 
We  denote with $q$ and $q'$ two quarks having different flavors, and 
similar notation holds for the antiquarks. 
Furthermore, the following convolutions of unpolarized parton
distributions are defined
\begin{eqnarray}
{\cal{F}}^{a  b}(x_1, x_2, \boldsymbol{q}_{T}^2) & \equiv & 
\int d^2\boldsymbol{p}_{1 T}\,d^2
\boldsymbol{p}_{2 T} 
\,\delta^2 (\boldsymbol{p}_{1 T} +\boldsymbol{p}_{2 T} -\boldsymbol{q}_{T})
  f_1^{a}(x_1, \boldsymbol p^2_{1 T}) f_1^{{b}} 
(x_2, \boldsymbol p^2_{2 T})\, , 
\label{eq:Fqq}
\end{eqnarray}
where a sum over all (anti)quark flavors is understood. Our results
for the terms ${\cal A}^{ab\to c d}$ in Eq.\ (\ref{eq:A}) are listed below, 
starting from  the ones corresponding to the (anti)quark induced processes,
\begin{eqnarray}
{\cal A}^{q  q' \to q q'} & = & a(y){\cal{F}}^{q q'} 
(x_1, x_2,\boldsymbol{q}_{T}^2) + a(1-y) {\cal{F}}^{q'q} 
(x_1, x_2,\boldsymbol{q}_{T}^2) \, ,
\label{eq:a1}
\end{eqnarray}
\begin{eqnarray}
{\cal A}^{q  \bar q' \to q \bar q' } & = &  
a(y) {\cal{F}}^{q \bar q'} 
(x_1, x_2,\boldsymbol{q}_{T}^2) + a(1-y) {\cal{F}}^{\bar q' q} 
(x_1, x_2,\boldsymbol{q}_{T}^2) \, ,
\end{eqnarray}
\begin{eqnarray}
{\cal A}^{q  q \to q q } & = &\frac{N^2-1}{2 N^2}\, y (1-y) \, \bigg [ 
\frac{1+(1-y)^2}{y^2} + \frac{1+y^2}{(1-y)^2} - 
\frac{2}{N} \frac{1}{y(1-y)} \bigg ] {\cal{F}}^{q q}(x_1, x_2, \boldsymbol{q}_{T}^2) \, , 
\label{eq:Aqqqq}
\end{eqnarray}
\begin{eqnarray}
{\cal A}^{q \bar{q} \to q \bar{q}} &= & b(y){\cal{F}}^{q \bar{q}}
(x_1, x_2, \boldsymbol{q}_{T}^2) + b(1-y) {\cal{F}}^{\bar{q} q}(x_1, x_2, \boldsymbol{q}_{T}^2)\,, 
\label{eq:Aqqbqqb}
\end{eqnarray}
\begin{eqnarray}
{\cal A}^{q \bar{q} \to q' \bar q'} &= & \frac{N^2-1}{2 N^2}\, y (1-y) \,
\left [y^2 + (1-y)^2 \right ]   \bigg [{\cal{F}}^{q \bar{q}}(x_1, x_2, \boldsymbol{q}_{T}^2)+ {\cal{F}}^{\bar q q}(x_1, x_2, \boldsymbol{q}_{T}^2) \bigg ]       \, , 
\end{eqnarray}
\begin{eqnarray}
{\cal A}^{q \bar{q} \to g g} & = & \frac{N^2-1}{N}\bigg (y^2 + (1-y)^2-\frac{1}{N^2} \bigg )\frac{ y^2 + (1-y)^2}{2} \bigg [{\cal{F}}^{q \bar{q}}(x_1, x_2, \boldsymbol{q}_{T}^2)+ {\cal{F}}^{\bar q q}(x_1, x_2, \boldsymbol{q}_{T}^2) \bigg ]\, ,
\end{eqnarray}
with $N$ being the number of colors and
\begin{eqnarray}
a(y) &=& \frac{N^2-1}{2 N^2} \,(1-y) \frac{1 + (1-y)^2}{y}\, , \qquad
b(y) \, = \,a(y) +\frac{N^2-1}{2 N^2}\, y (1-y) \,
\bigg [y^2 + (1-y)^2  +  \frac{2}{N} \frac{(1-y)^2}{y} \bigg ]\,.
\end{eqnarray}
Analogously, from the gluon induced processes, one has:
\begin{eqnarray}
{\cal A}^{q  g \to q g } & = & c(y) {\cal{F}}^{q g} 
(x_1, x_2,\boldsymbol{q}_{T}^2) + c(1-y){\cal{F}}^{g q} (x_1, x_2,\boldsymbol q_T^2)\, ,
\end{eqnarray}
\begin{eqnarray}
{\cal A}^{g g \to g g } &= & 4\, \frac{N^2}{N^2-1}\, \frac{(1-y (1-y))^3}{y (1-y)}{\cal{F}}^{g g} (x_1, x_2,\boldsymbol q_T^2)\, ,
\label{eq:Agg1}
\end{eqnarray}
\begin{eqnarray}
{\cal A}^{g g \to q \bar{q} } & =&  \frac{N}{N^2-1}\,  \bigg (y^2 + (1-y)^2 - \frac{1}{N^2}\bigg ) \,
\frac{y^2 + (1-y)^2}{2}\, {\cal{F}}^{g g} (x_1, x_2,\boldsymbol q_T^2)\, ,
\label{eq:Agg2}
\end{eqnarray} 
where
\begin{equation}
c(y) = \frac{1+(1-y)^2}{2}\,\bigg [\frac{1+(1-y)^2}{y}-\frac{y}{N^2}\bigg ]~.
\end{equation}
Agreement is found between the results given in the present subsection and the
explicit expressions of the partonic cross sections published, for example,
in
\cite{Owens:1977sj,Combridge:1977dm,Cutler:1977qm,Jaffe:1996ik,Bomhof:2006ra}. 
However, with respect to Ref.\ \cite{Lu:2008qu} we find agreement with
the expression for the unpolarized $q q $ production subprocess, but not
for the $q \bar{q}$ production subprocesses. In particular, we find 
differences as compared with their Eqs.\ (24) and (33). 

%%%%%%%%%%%%%%%%%%%%%%%%%%%%%%%%%%%%%%%%%%%%%%%%%%%%%%%%%%%%%%%%%%%%%%%%%%%%%%%
\subsection{The $\cos 2 ( \phi_T-\phi_\perp)$ angular distribution of the dijet}
%%%%%%%%%%%%%%%%%%%%%%%%%%%%%%%%%%%%%%%%%%%%%%%%%%%%%%%%%%%%%%%%%%%%%%%%%%%%%%%

In Ref.\ \cite{Lu:2008qu} it is shown that the subprocesses $qq\to qq$ and $q\bar{q}\to q\bar{q}$ contribute not only to the angular independent part of the 
cross section, according to Eqs.\ (\ref{eq:Aqqqq}) and  (\ref{eq:Aqqbqqb}), 
but also   
to an azimuthal asymmetry of the dijet
arising from the product of two T-odd 
functions, $h_1^{\perp q} h_1^{\perp q}$ or 
$h_1^{\perp q} h_1^{\perp \bar{q}}$. Such an asymmetry is similar to the one
calculated in the Drell-Yan \cite{Boer:1999mm} and in the photon-jet
production \cite{Boer:2007nd} processes. 
We refer to \cite{Boer:2007nd} for the details 
of the derivation and present here only our final results. In analogy to  
Eq.\ (\ref{eq:A}), we write
\begin{equation}
B(y, x_1, x_2, \boldsymbol{q}_{T}^2) = 
\sum_{a, b, c, d} {\cal B}^{a b\to c d}(y, x_1, x_2, \boldsymbol{q}_{T}^2)\, , 
\label{eq:B}
\end{equation}
with 
\begin{eqnarray}
{\cal B}^{q  q \to q q}& =&\frac{N^2-1}{N^3}\, y (1-y) 
{\cal{H}}^{q q }(x_1, x_2,\boldsymbol{q}_{T}^2) \, ,
\label{eq:Bqq}
\end{eqnarray}
\begin{eqnarray}
{\cal B}^{q  \bar q \to q \bar q}& =& d(y) {\cal{H}}^{q \bar{q}}
(x_1, x_2,\boldsymbol{q}_{T}^2)
+ d(1-y) {\cal{H}}^{\bar q q }(x_1, x_2,\boldsymbol{q}_{T}^2) \, ,
\label{eq:Bqqb} 
\end{eqnarray}
and 
\begin{eqnarray}
d(y)& =& \frac{N^2-1}{N^3}\, y (1-y)^2\,(1 + N y)~.
\end{eqnarray} 
The following convolution of (transversely polarized) quark and antiquark 
distributions has been introduced 
\begin{eqnarray}
\boldsymbol{q}_T^2\,{\cal{H}}^{q \bar q}(x_1, x_2, \boldsymbol{q}_{T}^2) 
& \equiv &
\frac{1}{M_1 M_2}\sum_{\rm flavors}  \int d^2\boldsymbol{p}_{1 T}\,d^2
\boldsymbol{p}_{2 T} 
\,\delta^2 (\boldsymbol{p}_{1 T} +\boldsymbol{p}_{2 T} -\boldsymbol{q}_{T})
\nonumber \\
&& \qquad \qquad\qquad \mbox{} \times 
\left(2 (\boldsymbol{\hat{h}} \cdot \boldsymbol{p}_{1 T})
(\boldsymbol{\hat{h}} \cdot \boldsymbol{p}_{2 T}) 
- ( \boldsymbol{p}_{1 T}\cdot \boldsymbol{p}_{2 T}) \right)
\ h_1^{\perp q}(x_1, \boldsymbol p^2_{1 T}) 
h_1^{\perp \bar q} (x_2, \boldsymbol p^2_{2 T})\, ,
\label{eq:Hqq}
\end{eqnarray}
with $\hat{\boldsymbol{h}} \equiv \boldsymbol{q}_{T}/|\boldsymbol{q}_{T}|$, 
and a  similar definition holds for ${\cal H}^{q q}$ upon 
replacement of $\bar q\to q$ in Eq.\ (\ref{eq:Hqq}).  
The small-$\boldsymbol{q}_T$ behavior of ${\cal H}$ is regular provided
the integrations over 
$\boldsymbol{p}_T^4\,h_1^{\perp q}(x,\boldsymbol{p}_T^2)$ converge.
In addition to Eqs.\ (\ref{eq:Bqq}) and (\ref{eq:Bqqb}), we find that the 
subprocesses 
$q \bar{q}\to g g$ and  $q \bar{q}\to q' \bar{q}'$, not considered in 
\cite{Lu:2008qu}, also show a $\cos 2 (\phi_T-\phi_\perp)$ angular dependence,
leading respectively to 
\begin{eqnarray}
{\cal B}^{q \bar{q} \to g g }& = & \frac{N^2-1}{N}\bigg (y^2 + (1-y)^2-\frac{1}{N^2} \bigg ) y (1-y) 
\bigg [ {\cal{H}}^{q \bar{q}}(x_1, x_2,\boldsymbol{q}_{T}^2) +
 {\cal{H}}^{\bar q q}(x_1, x_2,\boldsymbol{q}_{T}^2) \bigg ]\, , 
\end{eqnarray}
and 
\begin{eqnarray}
{\cal B}^{q  \bar q \to q' \bar q' } & = & \frac{N^2-1}{N^2}\, y^2 (1-y)^2 \bigg [ {\cal{H}}^{q \bar{q}}(x_1, x_2,\boldsymbol{q}_{T}^2) +
 {\cal{H}}^{\bar q q}(x_1, x_2,\boldsymbol{q}_{T}^2) \bigg ]~.
\end{eqnarray}
Agreement is found between the results given in the present subsection
  and the explicit expressions of the polarized partonic cross sections 
published 
in \cite{Jaffe:1996ik,Bomhof:2006ra,Ji:1992ev,Soffer:2002tf}.
For the polarized $q q$ production subprocess we find 
agreement with Ref.\ \cite{Lu:2008qu}, but again not for the $q \bar{q}$ 
production subprocesses (in particular, we find a difference compared to their
Eq.\ (26)).

%%%%%%%%%%%%%%%%%%%%%%%%%%%%%%%%%%%%%%%%%%%%%%%%%%%%%%%%%%%%%%%%%%%%%%%%%%%%%%%
\subsection{The $\cos 4(\phi_T-\phi_\perp) $ angular distribution of the dijet }
%%%%%%%%%%%%%%%%%%%%%%%%%%%%%%%%%%%%%%%%%%%%%%%%%%%%%%%%%%%%%%%%%%%%%%%%%%%%%%%
The  $\cos 4( \phi_T-\phi_\perp) $ angular distribution of the dijet
is related to
the presence of linearly polarized gluons in  unpolarized  hadrons. 
This being a new 
result of the present paper, its derivation will be discussed in some more 
detail.
The gluon-gluon induced part of the reaction under study, to lowest order in 
pQCD, 
is described in terms of the partonic two-to-two subprocesses
\begin{equation}
g(p_1) + {g}(p_2) \rightarrow g (K_1) + g(K_2)\, ,\qquad {\rm and} \qquad
g(p_1) + {g}(p_2) \rightarrow q (K_1) + \bar{q}(K_2)~.
\label{eq:sub}
\end{equation} 
The corresponding cross sections are given by 
\begin{eqnarray}
\frac{d \sigma^{g g\to g g}}
{d\eta_1\, \, d\eta_2 \,d^2 \boldsymbol K_{\perp}\, d^2 \boldsymbol q_{T}}
&  =   & \frac{\alpha_s^2} {s   \boldsymbol K_{\perp} ^2 }  
\bigg [ {\cal A}^{g g \to g g }(y, x_1, x_2, \boldsymbol q_T^2) + \int d^2\boldsymbol{p}_{1 T}\,d^2
\boldsymbol{p}_{2 T} 
\,\delta^2 (\boldsymbol{p}_{1 T} +\boldsymbol{p}_{2 T} -\boldsymbol{q}_{T}) \nonumber  \\
& & \quad \times \frac{N^2}{N^2-1}\,y (1-y) (1-y (1-y))\,{\cal P}^{g g}
(\boldsymbol p_{1 T},\boldsymbol p_{2 T}, \boldsymbol K_{1 \perp},\boldsymbol K_{2 \perp}) \bigg ]\, ,
\label{eq:cs}
\end{eqnarray}
and
\begin{eqnarray}
\frac{d\sigma^{g g\to q\bar{q}}}
{d\eta_1\, d\eta_2 \,d^2 \boldsymbol K_{\perp}\, d^2 \boldsymbol q_{T} }
&  = &\frac{\alpha_s^2} {s \boldsymbol K_{\perp} ^2 }\, 
\bigg [  {\cal A}^{g g \to q \bar q}(y, x_1, x_2, \boldsymbol q_T^2)
- \int d^2\boldsymbol{p}_{1 T}\,d^2
\boldsymbol{p}_{2 T} 
\,\delta^2 (\boldsymbol{p}_{1 T} +\boldsymbol{p}_{2 T} -\boldsymbol{q}_{T}) \nonumber \\
&&\quad \times \frac{N}{N^2-1}\,\frac{y (1-y)}{4} \, \bigg (y^2 + (1-y)^2 - \frac{1}{N^2}\bigg ) \, {\cal{P}}^{g g }
(\boldsymbol p_{1 T},\boldsymbol p_{2 T}, \boldsymbol K_{1 \perp},\boldsymbol K_{2 \perp}) \bigg ]\, ,
\label{eq:csphi}
\end{eqnarray} 
where 
\begin{eqnarray}
{\cal{P}}^{g g }
(\boldsymbol p_{1 T},\boldsymbol p_{2 T}, \boldsymbol K_{1 \perp},\boldsymbol K_{2 \perp})& = &\bigg [
-\boldsymbol{p}_{1 T}^2\boldsymbol{p}_{2 T}^2 + \frac{2}{\boldsymbol{K}_{\perp} ^4} \, \bigg (
(\boldsymbol{K}_{1\perp}\cdot \boldsymbol K_{2 \perp})
(\boldsymbol{p}_{1 T}\cdot\boldsymbol{p}_{2 T}) - 
{(\boldsymbol{K}_{1 \perp}\cdot \boldsymbol {p}_{1 T}) 
(\boldsymbol{K}_{2 \perp} \cdot \boldsymbol{p}_{2 T})}\nonumber  \\
&&~~~~~~ -(\boldsymbol{K}_{1 \perp}\cdot \boldsymbol {p}_{2 T}) 
(\boldsymbol{K}_{2 \perp} \cdot \boldsymbol{p}_{1 T}) 
\bigg )^2 \bigg ] \frac{1}{M_1^2 M_2^2} \,h_1^{\perp g}(x_1, \boldsymbol p^2_{1 T})\, h_1^{\perp g} 
(x_2, \boldsymbol p^2_{2 T})~. 
\end{eqnarray}
The functions $ {\cal A}^{g g \to g g }$ and $ {\cal A}^{g g \to q \bar q}$, 
given in Eqs.\ (\ref{eq:Agg1}) and (\ref{eq:Agg2}), contain the convolution
of unpolarized gluon distribution functions ${\cal F}^{g g}$ defined in 
Eq.\ (\ref{eq:Fqq}). 
In order to show that the two cross sections in Eqs.\ (\ref{eq:cs})
and  (\ref{eq:csphi}) can be 
 written  in  the same form as Eq.\ (\ref{eq:cso}), we introduce the functions
\begin{eqnarray}
\boldsymbol{q}_T^4\,{\cal{I}}^{g g }(x_1, x_2, \boldsymbol{q}_T^2 ) 
& \equiv & 
\frac{1}{M_1^2 M_2^2}\int d^2\boldsymbol{p}_{1 T}\,d^2
\boldsymbol{p}_{2 T} 
\,\delta^2 (\boldsymbol{p}_{1 T} +\boldsymbol{p}_{2 T} -\boldsymbol{q}_{T}) 
\nonumber \\
&& \qquad \qquad\qquad \mbox{} \times 
\bigg (2 (\hat{\boldsymbol{h}} \cdot \boldsymbol{p}_{1 T})
( \hat{\boldsymbol{h}} \cdot \boldsymbol{p}_{2 T}) 
- (\boldsymbol{p}_{1 T} \cdot\boldsymbol{p}_{2 T}) \bigg )^2
\ h_1^{\perp g}(x_1, \boldsymbol p^2_{1 T}) h_1^{\perp g} 
(x_2, \boldsymbol p^2_{2 T}) \, ,
\label{eq:I}
\end{eqnarray}
where  again $\hat{\boldsymbol{h}} \equiv \boldsymbol{q}_{T}/|\boldsymbol{q}_{T}|$, and
\begin{equation}
\boldsymbol{q}_T^4\,{{\cal{L}}^{g g }(x_1, x_2,\boldsymbol{q}_T^2)} 
\equiv \frac{1}{M_1^2 M_2^2} 
\, \int d^2\boldsymbol{p}_{1 T}\,d^2
\boldsymbol{p}_{2 T} 
\,\delta^2 (\boldsymbol{p}_{1 T} +\boldsymbol{p}_{2 T} -\boldsymbol{q}_{T}) \boldsymbol p^2_{1 T} \boldsymbol p^2_{2 T}
 h_1^{\perp g}(x_1, \boldsymbol p^2_{1 T}) h_1^{\perp g} 
(x_2, \boldsymbol p^2_{2 T})~.
\label{eq:L} 
\end{equation}
The small-$\boldsymbol{q}_T$ behavior of ${\cal I}$ and ${\cal L}$ are regular 
provided the integrations over 
$\boldsymbol{p}_T^8\,h_1^{\perp g}(x,\boldsymbol{p}_T^2)$ converge.
Hence we have
%%%%%%%%%%%%%%%%%%%%%%%%%%%%%%%%%%%%%%%%%%%%%%%%%%%%%%%%%%%%%%%%%%%%%%%%%%%%%%%
%%%%%%%%%%%%%%%%%%%%%%%%%%%%%%%%%%%%%%%%%%%%%%%%%%%%%%%%%%%%%%%%%%%%%%%%%%%%%%%
\begin{eqnarray}
&&\int d^2\boldsymbol{p}_{1 T}\,d^2
\boldsymbol{p}_{2 T}\,\delta^2 (\boldsymbol{p}_{1 T} +\boldsymbol{p}_{2 T} -\boldsymbol{q}_{T}) {\cal P}^{g g } \nonumber \\
&& \qquad \mbox{}
= \boldsymbol{q}_T^4\sum_{i, j, l, m =1}^2 \,
\frac{\boldsymbol{K}_{1 \perp}^{\{ i}\boldsymbol{K}_{2 \perp}^{j\}}
\boldsymbol{K}_{1 \perp}^{\{ l}\boldsymbol{K}_{2 \perp}^{m\}}}
{4  \,\boldsymbol{K}_{\perp}^4 }
\bigg [2 (\delta^{i l} \delta^{j m} -\delta^{i j} \delta^{l m} + \delta^{i m} 
\delta^{j l}) ({\cal{L}}^{g g }-{\cal{I}}^{g g })
\nonumber \\
&& \qquad\quad \mbox{}
-\delta^{i l}\delta^{j m} {\cal{L}}^{g g }+ 2 \bigg (\frac{\boldsymbol{q}_{T}^{\{i}\boldsymbol{q}_{T}^{j\}}}
{\boldsymbol{q}_{T}^2} - \delta^{i j} \bigg )\bigg (\frac{\boldsymbol{q}_{T}^{\{l}\boldsymbol{q}_{T}^{m\}}}
{\boldsymbol{q}_{T}^2} - \delta^{l m} \bigg )
(2 {\cal{I}}^{g g }- {\cal{L}}^{g g }) \bigg ]  \nonumber \\
&& \qquad \mbox{}
= \boldsymbol{q}_T^4 \,\cos 2 (2\phi_T-\phi_1-\phi_2)(2 {\cal I}^{g g }-{\cal L}^{g g })\, .
\label{eq:csphi2}
\end{eqnarray}
The difference between the angular dependence
$\cos 2 (2\phi_T-\phi_1-\phi_2)= \cos 4(\phi_T-\phi_j)$ and 
$\cos 4 (\phi_T-\phi_\perp)$ is
of order $\boldsymbol{q}_T^2/\boldsymbol{K}_\perp^2$ (cf.\
Appendix~\ref{appendixA}). 
Substituting Eq.\ (\ref{eq:csphi2}) into Eqs.\ (\ref{eq:cs}) and
(\ref{eq:csphi}), and defining  $d\sigma^{g g } 
\equiv d\sigma^{g g \to gg} + d\sigma^{gg\to q \bar q}$, we finally obtain  
\begin{eqnarray}
\frac{d\sigma^{g g}}
{d\eta_1\, d\eta_2 \,d^2 \boldsymbol K_{\perp}\,d^2 \boldsymbol q_{T} } 
& = & \frac{\alpha_s^2}{s  \boldsymbol K_{\perp}^2}
\bigg [ {\cal A}^{g g }(y, x_1, x_2, \boldsymbol{q}_T^2) + 
{\cal C}^{g g} (y, x_1, x_2, \boldsymbol{q}_T^2)
\,\boldsymbol{q}_T^4\,\cos 4 (\phi_T-\phi_\perp)  
\bigg ]\, , \nonumber \\
\label{eq:cross}
\end{eqnarray}
where
${\cal A}^{g g}\equiv {\cal A}^{g g \to g g } +
 {\cal A}^{g g \to q \bar{q} }$, 
${\cal C}^{g g}\equiv {\cal C}^{g g \to g g } +
 {\cal C}^{g g \to q \bar{q} }$, with
\begin{eqnarray}
{\cal C}^{g g \to g g } = \frac{N^2}{N^2-1}\,y (1-y) (1-y (1-y))\,
\bigg [2 {\cal{I}}^{g g }(x_1, x_2,\boldsymbol{q}_T^2) - {\cal{L}}^{g g }(x_1, x_2,\boldsymbol{q}_T^2)\bigg ]\, , 
\label{eq:Cgg}
\end{eqnarray}
and 
\begin{eqnarray}
{\cal C}^{g g \to q \bar q} = -\frac{N}{N^2-1}\,\frac{y (1-y)}{4} \, \bigg (y^2 + (1-y)^2 - \frac{1}{N^2}\bigg ) \, \bigg [2 {\cal{I}}^{g g }(x_1, x_2,\boldsymbol{q}_T^2) - {\cal{L}}^{g g }(x_1, x_2,\boldsymbol{q}_T^2)\bigg ]~.
\label{eq:Cqqb}
\end{eqnarray}

It turns out that the two subprocesses $g g\to g g$ and  $g g\to q \bar q$ are 
the only ones that determine the $\cos 4(\phi_T-\phi_\perp)$ dependence of the 
cross section. Therefore in Eq.\ (\ref{eq:cso})
\begin{eqnarray}
C(y, x_1, x_2, \boldsymbol q_T^2) & = & {\cal C}^{g g} =  {\cal C}^{g g \to g g }+ {\cal C}^{g g \to q \bar q} \, ,
\label{eq:Cf}
\end{eqnarray}
which, together with Eqs.\ (\ref{eq:I})-(\ref{eq:L}) and 
Eqs.\ (\ref{eq:Cgg})-(\ref{eq:Cqqb}), leads to
\begin{equation}
C  = \frac{N}{N^2-1}\,y (1-y) \bigg [ N (1-y (1-y)) -\frac{1}{4} 
\bigg (y^2 + (1-y)^2-\frac{1}{N^2} \bigg )\bigg ] \bigg [ 2 {\cal I}^{g g } (x_1, x_2, \boldsymbol{q}_T^2) -{\cal L}^{g g } (x_1, x_2, 
\boldsymbol{q}_T^2) \bigg ]\, ,
\label{eq:Cff}
\end{equation}
showing how the azimuthal asymmetry under investigation
is related to the T-even, spin and transverse momentum 
dependent parton distribution function $h_1^{\perp g}(x, \boldsymbol p_{T}^2)$.

%%%%%%%%%%%%%%%%%%%%%%%%%%%%%%%%%%%%%%%%%%%%%%%%%%%%%%%%%%%%%%%%%%%%%%%%%%%%%%
\section{\label{sec:imbalance}
Dijet imbalance distributions}
%%%%%%%%%%%%%%%%%%%%%%%%%%%%%%%%%%%%%%%%%%%%%%%%%%%%%%%%%%%%%%%%%%%%%%%%%%%%%%

In this section we study the cross section for the process
$h_1\, h_2 \to {\rm jet} \, {\rm jet}\, X$ in terms of the total transverse 
energy $E_T$ and  the dijet imbalance $\delta\phi \equiv \phi_1 -\phi_2 -\pi$, 
which are the kinematic variables commonly used in the experiments.
The dijet imbalance angle describes the deviation of the two jets from a 
back-to-back configuration (see Fig.\ \ref{transverseplane} in 
Appendix~\ref{appendixA}). 

The transverse energy is the sum of the transverse energies of the two
jets, $E_T= |\boldsymbol K_{1 \perp}|+ |\boldsymbol K_{2 \perp}|$, and 
the difference is defined as 
$\Delta K_{\perp} = |\boldsymbol K_{1 \perp}| - |\boldsymbol K_{2 \perp}|$.
In our basic expression for the cross section in Eq.\ (\ref{eq:cso}) we have
traded $\boldsymbol{K}_{1\perp}$ and $\boldsymbol{K}_{2\perp}$ for 
$\boldsymbol{q}_T$ and $\boldsymbol{K}_\perp$, but we can also trade
the variables ($|\boldsymbol K_{1 \perp}|$,
$|\boldsymbol K_{2 \perp}|$) for ($E_T$, $\Delta K_{\perp}$) and
($\phi_1$, $\phi_2$) for ($\phi_j$, $\delta\phi$). 
We find in the back-to-back approximation
\begin{eqnarray}
&&
\boldsymbol q_T^2 = 
\Delta K_\perp^2\,\cos^2\left (\frac {\delta\phi}{2}\right )
+ E_T^2 \sin^2\left (\frac{\delta\phi}{2}\right ) 
\approx 
\Delta K_\perp^2 +
E_T^2 \sin^2\left (\frac{\delta\phi}{2}\right ) \, , 
\label{eq:qt2}
\\&&
4\,\boldsymbol{K}_\perp^2 = 
E_T^2 \cos^2\left (\frac{\delta\phi}{2}\right ) + 
\Delta K_\perp^2\,\sin^2\left (\frac {\delta\phi}{2}\right )
\approx E_T^2\ .
\end{eqnarray}
In the first expression we cannot drop the term proportional 
to $\Delta K_\perp^2$ because it is not a good approximation for 
$\delta \phi \approx 0$, which is most relevant.
Note also that this implies $\boldsymbol q_T^2 \geq \Delta K_\perp^2$, i.e.\ 
$\Delta K_\perp^2$ sets a lower
bound on the $\boldsymbol q_T^2$ values probed, which may be very relevant if 
the functions $A,B,C$ are steeply falling functions with increasing $\boldsymbol q_T^2$. 

The cross section in Eq.\ (\ref{eq:cso}) rewritten yields 
\begin{equation}
\frac{d\sigma^{h_1 h_2 \rightarrow {\rm jet} \,{\rm jet} \, X}}{d\eta_1 d\eta_2 d E_T d\Delta K_\perp d\phi_j d\delta\phi} = 
\frac{\alpha_s^2}{2 s} \bigg [ A(\boldsymbol{q}_T^2) 
+  {B}(\boldsymbol{q}_T^2) \, \boldsymbol{q}_T^2 \,\cos 2 (\phi_T-\phi_j) +
 {C}(\boldsymbol{q}_T^2) \, \boldsymbol{q}_T^4 \, 
\cos 4 (\phi_T-\phi_j) \bigg ]\, ,
\label{xsdeltaphi}
\end{equation}
with $\boldsymbol q_T^2$ given in the unapproximated first part of 
Eq.\ (\ref{eq:qt2}) and 
\begin{eqnarray}
\boldsymbol q_T^2\,\cos 2 (\phi_T-\phi_j) 
& = & \Delta K_\perp^2\,
\cos^2\left (\frac {\delta\phi}{2}\right )
- E_T^2 \sin^2\left (\frac{\delta\phi}{2}\right ) \, , \\
\boldsymbol q_T^4\,\cos 4 (\phi_T-\phi_j) & =& 
\left[ E_T^2 \sin^2\left (\frac{\delta\phi}{2}\right ) - 
\Delta K_\perp^2\,\cos^2\left (\frac {\delta\phi}{2}\right )\right]^2 - 
E_T^2 \,\Delta K_\perp^2 \, \sin^2 (\delta\phi)\, . 
\end{eqnarray}
In this way we have arrived at an expression that is amenable 
to phenomenological studies, approximating 
$ |\boldsymbol K_{1 \perp}|\approx  |\boldsymbol K_{2 \perp}| 
\approx E_T/2$ only in places where the difference is negligible for
all values of $\delta \phi$. 

For $\Delta K_\perp=0$ we obtain: 
\begin{equation}
\frac{d\sigma^{h_1 h_2 \rightarrow {\rm jet} \,{\rm jet} \, X}}{d\eta_1
  d\eta_2 d E_T d\Delta K_\perp d\phi_j d\delta\phi} =   
\frac{\alpha_s^2}{2 s} \bigg [ A(\boldsymbol{q}_T^2) 
-  {B}(\boldsymbol{q}_T^2) \boldsymbol q_T^2 +
 {C}(\boldsymbol{q}_T^2) \boldsymbol{q}_T^4 \bigg ]\, ,
\label{xsdeltaphiideal}
\end{equation}
with in that case exactly 
$\boldsymbol q_T^2 = E_T^2 \sin^2\left(\delta\phi/2\right)$,
and in essence recovering Eq.\ (\ref{schematicxs}) (the sign in
  front of $B$ is just a matter of definition). 

To illustrate the effect of nonzero $B$ and $C$ terms, we will make a
Gaussian Ansatz for these functions of $\boldsymbol{q}_T^2$. We will take: 
\begin{equation}
A(\boldsymbol{q}_T^2) = \frac{R_A^2}{\pi} \exp(-\boldsymbol{q}_T^2
R_A^2), \qquad 
B(\boldsymbol{q}_T^2) = \frac{R_B^4}{c\pi} \exp(-\boldsymbol{q}_T^2
R_B^2), \qquad
C(\boldsymbol{q}_T^2) = \frac{R_C^6}{2 c^2\pi} \exp(-\boldsymbol{q}_T^2
R_C^2),
\end{equation}
normalized such that 
\begin{equation}
\int d^2 \boldsymbol{q}_T A(\boldsymbol{q}_T^2) = 1, \qquad 
\int d^2 \boldsymbol{q}_T \boldsymbol{q}_T^2 B(\boldsymbol{q}_T^2) =
1/c,\qquad
\int d^2 \boldsymbol{q}_T \boldsymbol{q}_T^4 C(\boldsymbol{q}_T^2) =
1/c^2.
\end{equation}
Fig.\ \ref{illustration} shows a plot of the cross section in Eq.\
(\ref{xsdeltaphi}) as a function of $\delta \phi$ for the arbitrary,
but perhaps realistic choices
$|\boldsymbol K_{1 \perp}|=30$ GeV,
$|\boldsymbol K_{2 \perp}|=31$ GeV, $R_A=0.5$ GeV$^{-2}$, 
$R_B=2R_A, R_C=3R_A$ and $c=3$. 
\begin{figure}[htb]
\epsfig{figure=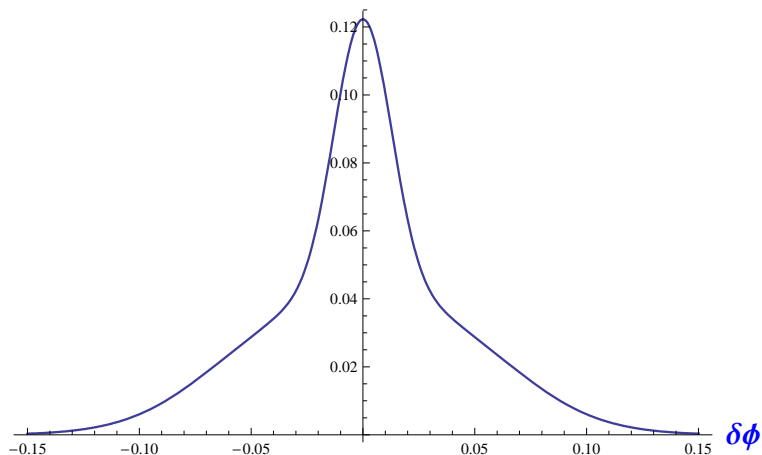,width=10cm} 
\caption{\label{illustration} An illustration of the effect of
  sizeable $B$ and $C$ terms on the $\delta \phi$ distribution of the 
cross section in Eq.\ (\ref{xsdeltaphi}).}
\end{figure}
For smaller $\Delta K_\perp$ the shoulders become more pronounced, but 
already 
for $\Delta K_\perp=2$ GeV the shoulders are hardly distinguishable 
anymore. 
In general, $B$ and $C$ have to be significant in size and broad enough
to generate an observable effect, i.e.\ for 
the $\delta \phi$ distribution to deviate visibly from a Gaussian distribution.

Although the $B$ and $C$ terms were not considered before in
experimental analyses of dijet imbalance measurements in hadronic
collisions, there is experimental data available that has some bearing
on the size of $B$ compared to $A$. It comes from the measurement of
the violation of the Lam-Tung relation in the Drell-Yan process. As
shown in Ref.\ \cite{Boer:1999mm}, this violation $\kappa$ is given by
 the ratio $\boldsymbol{q}_T^2\,{\cal{H}}^{q \bar q}/{\cal{F}}^{q \bar q}$
 (Eq.\ (\ref{eq:Hqq}) divided by the angular averaged result in
 Eq.~(\ref{eq:Fqq})) but with the
 sums over flavors weighted with a factor $e_q^2$, the quark charge
 squared. This has the effect of emphasizing the contribution
 from up quarks. In the present two-jet production case, the ratio
$\boldsymbol{q}_T^2\,B/A$ in the midrapidity region ($\eta_1 \approx
\eta_2 \approx 0$) and for large $N$ can be approximated by
$\boldsymbol{q}_T^2\,{\cal{H}}^{q \bar q}/{\cal{F}}^{q \bar q}$. The
size of $\kappa$ in Drell-Yan may thus be expected to give some
indication of the size of $\boldsymbol{q}_T^2\,B/A$. The violation of
the Lam-Tung relation in Drell-Yan has recently been measured in $p p$
and $p d$ collisions \cite{Zhu:2008sj}. It is consistent with no
violation, but with sizeable errors. Small violation would be in line
with the expectation that $h_1^\perp$ for antiquarks inside a proton
is considerably smaller than for quarks. For $p \bar p$ one however
expects a large violation, as observed in $\pi p$ collisions
\cite{Falciano:1986wk,Guanziroli:1987rp,Conway:1989fs}. So the effect
of a nonzero $h_1^{\perp}$ for {\em quarks} may be mostly relevant for jet
broadening studies in $p \bar p$~\cite{Abe:1991dma}.

%%%%%%%%%%%%%%%%%%%%%%%%%%%%%%%%%%%%%%%%%%%%%%%%%%%%%%%%%%%%%%%%%%%%%%%%%%%%%%
\section{Weighted cross sections \label{sec:weightedxs}}
%%%%%%%%%%%%%%%%%%%%%%%%%%%%%%%%%%%%%%%%%%%%%%%%%%%%%%%%%%%%%%%%%%%%%%%%%%%%%%

Apart from the fact that nonzero $h_1^\perp$ functions for quarks
and gluons modify the $\delta \phi$ distribution and hence affect
the extraction of the average initial parton transverse momentum
from this dijet imbalance distribution, it would in principle be of
interest to extract these functions themselves from it.
Therefore, the question arises whether one can
project out the $B$ and $C$ terms separately. In Ref.\ 
\cite{Lu:2008qu} this is discussed for $B$ only, but there are some
problems with the proposed method. It was suggested that $\langle \cos
\delta \phi \rangle$, i.e.\ the cross section integrated over
$\delta \phi$ weighted with an additional factor of $\cos \delta
\phi$, projects out a contribution from $h_1^{\perp \, q}$
exclusively\footnote{In Ref.\ \cite{Lu:2008qu} actually $\langle
P_\perp^2/M^2 \cos \delta \phi \rangle $ was considered, where
$P_\perp \approx |\boldsymbol K_{1\perp}| \approx |\boldsymbol
K_{2 \perp}|$, despite the fact that $\boldsymbol K_{2 \perp}$ was
integrated over. The factor $P_\perp^2/M^2$ artificially enhances the
weighted asymmetry if not divided by $\langle P_\perp^2/M^2
\rangle$.}. 
However, our result in Eq.\ (\ref{xsdeltaphi}) shows
that $\langle \cos \delta \phi \rangle$ does not project out $B$ nor
a part of $B$ exclusively, not even
in the idealized case when $|\boldsymbol K_{1 \perp}| = |\boldsymbol
K_{2 \perp}|$, as can be seen from Eq.\ (\ref{xsdeltaphiideal}).
To see the appropriate weighting, 
we return to the form in Eq.\ (\ref{eq:cso}) and
note that $B$ is projected out by 
\begin{equation}
\big \langle \cos 2 (\phi_T-\phi_\perp)\big \rangle  \equiv  
\int \frac{d\phi_T}{2\pi}
\ \cos 2 (\phi_T-\phi_\perp)
\,{\frac{d\sigma^{h_1 h_2\to {\rm jet} \,{\rm jet}\, X}}
{d\eta_1\, d\eta_2 \,d^2 \boldsymbol K_{\perp}
\,d^2 \boldsymbol q_{T} }}
=  \frac{1}{2}\, \frac{\alpha_s^2}{s \boldsymbol K_{ \perp} ^2 }\, 
\boldsymbol{q}_T^2\, B(y, x_1, x_2, \boldsymbol{q}_T^2)~ .
\end{equation}
Integrating over the length of $\boldsymbol{q}_T$ gives with 
possible inclusion of additional weighting with powers of 
$\boldsymbol{q}_T^2$, 
\begin{equation}
\pi\int d\boldsymbol{q}_T^2 \ 
\ \left(\frac{\boldsymbol{q}_T^2}{M_1M_2}\right)^M
\,\big \langle \cos 2 (\phi_T-\phi_\perp)\big \rangle  
= \int d^2\boldsymbol{q}_T\ 
\ \left(\frac{\boldsymbol{q}_T^2}{M_1M_2}\right)^M
\,\cos 2 (\phi_T-\phi_\perp)
\,{\frac{d\sigma^{h_1 h_2\to {\rm jet} \,{\rm jet}\, X}}
{d\eta_1\, d\eta_2 \,d^2 \boldsymbol K_{\perp}
\,d^2 \boldsymbol q_{T} }}\, ,
\end{equation}
in which we get for $M = 1$ the factorized result
\begin{equation}
\pi\int d\boldsymbol{q}_T^2 
\ \left(\frac{\boldsymbol{q}_T^2}{M_1M_2}\right)
\,\boldsymbol{q}_T^2\, {\cal{H}}^{q q }(x_1, x_2, \boldsymbol{q}_T^2)
= 8\sum_{{\rm flavors}}\,h_1^{\perp q(1)}(x_1)\,h_1^{\perp q(1)}(x_2),
\end{equation}
in the $B$ contributions.
For $M = 0$, the expression does not deconvolute. In that case 
usable, but model dependent expressions may be obtained by making a 
Gaussian Ansatz for the transverse momentum shape of
$h_1^q$. For the type of convolution that appears in $B$
  this has been done in the literature, see for instance Ref.\
  \cite{Boer:1997mf}. 

Next, we will analyze in some more detail 
the weighted asymmetry that projects out $C$. 
A measurement of the weighted cross section 
\begin{eqnarray}
\big \langle \cos 4 (\phi_T-\phi_\perp )\big \rangle  \equiv  
\int \frac{d\phi_T}{2\pi}
\ \cos 4 (\phi_T-\phi_\perp)
\, {\frac{d\sigma^{h_1 h_2\to {\rm jet} \,{\rm jet}\, X}}
{d\eta_1\, d\eta_2 \,d^2 \boldsymbol K_{\perp}\, d^2 \boldsymbol q_{T} }}
=  \frac{1}{2}\, \frac{\alpha_s^2}{s \boldsymbol K_{\perp} ^2 }\, 
\boldsymbol{q}_T^4\, C(y, x_1, x_2, \boldsymbol{q}_T^2)\, ,
\label{eq:weight}
\end{eqnarray}
with $C$ given in Eq.\ (\ref{eq:Cff}), would give access to the linearly 
polarized gluon distribution of a hadron. 
After integration over the length of $\boldsymbol{q}_T$ with 
possible inclusion of additional weighting with $\boldsymbol{q}_T^2$,
we obtain
\begin{equation}
\pi\int d\boldsymbol{q}_T^2 \ 
\ \left(\frac{\boldsymbol{q}_T^2}{M_1M_2}\right)^M
\,\big \langle \cos 4 (\phi_T-\phi_\perp)\big \rangle  
= \int d^2\boldsymbol{q}_T\ 
\ \left(\frac{\boldsymbol{q}_T^2}{M_1M_2}\right)^M
\,\cos 4 (\phi_T-\phi_\perp)
\,{\frac{d\sigma^{h_1 h_2\to {\rm jet} \,{\rm jet}\, X}}
{d\eta_1\, d\eta_2 \,d^2 \boldsymbol K_{\perp}
\,d^2 \boldsymbol q_{T} }} \, .
\label{eq:weightq}
\end{equation}
In this case we get for $M = 2$ 
the deconvoluted result
\begin{equation}
\pi\int d\boldsymbol{q}_T^2 
\ \left(\frac{\boldsymbol{q}_T^2}{M_1M_2}\right)^2
\,\boldsymbol{q}_T^4\, \left(2\,{\cal I}^{g g} -{\cal L}^{g g}\right)
= 96\,h_1^{\perp g(2)}(x_1)\,h_1^{\perp g(2)}(x_2),
\end{equation}
in the $C$ contributions.
In order to evaluate the integral in Eq.\ (\ref{eq:weight}) without weights
or study the explicitly $\boldsymbol{q}_T^2$-dependence, one can 
employ a Gaussian model for $h_1^{\perp g}$, of which the easiest
choice has a factorized $x$ and $\boldsymbol{p}_T$
dependence, that is, neglecting the dependence on the factorization scale, 
\begin{eqnarray}
&&h_1^{\perp g}(x, \boldsymbol p_T^2) = 
\frac{R_h^2}{\pi}\,h_1^{\perp g}(x) \,e^{-R_h^2\boldsymbol p_T^2}\, ,  
\\ &&
h_1^{\perp g (n)}(x) = 
\int  d^2 \boldsymbol{p}_T\, 
\left(\frac{\boldsymbol{p}_T^2}{2M^2}\right)^n 
\,h_1^{\perp g }(x, \boldsymbol{p}_T^2) 
= \frac{n!}{(2\,M^2\,R_h^2)^n}\,h_1^{\perp g}(x),  
\end{eqnarray}
where $R_h$ is a size parameter related to the average partonic 
$\boldsymbol p_T^2$ by the relation  
$R_h^2 = 1/\langle \boldsymbol p_T^2 \rangle $. 
For incoming (anti)protons, $R_p = R_{\bar{p}}\equiv R$, so one has
\begin{eqnarray}
 \int d^2 \boldsymbol{q}_T \,\boldsymbol{q}_T^4\, 
(2 {\cal I}^{g g } -{\cal L}^{g g}) & = & 
\frac{1}{M_1^2 M_2^2}\int d^2\boldsymbol{q}_{T}\ d^2\boldsymbol{p}_{1 T}
\,d^2\boldsymbol{p}_{2 T} 
\,\delta^2 (\boldsymbol{p}_{1 T} +\boldsymbol{p}_{2 T} -\boldsymbol{q}_{T}) 
\bigg [ 2 \bigg (2 (\hat{\boldsymbol{h}} \cdot \boldsymbol{p}_{1 T})( \hat{\boldsymbol{h}} \cdot \boldsymbol{p}_{2 T}) - (\boldsymbol{p}_{1 T} \cdot\boldsymbol{p}_{2 T}) \bigg )^2
\nonumber \\
&& \qquad \qquad - \boldsymbol p_{1 T}^2 \boldsymbol p_{2 T}^2 \bigg ] \,
\frac{R^4}{\pi^2}\,h_1^{\perp g}(x_1) 
h_1^{\perp g}(x_2) e^{-R^2 (\boldsymbol p_{1 T}^2 + \boldsymbol p_{2 T}^2)}~.
\label{eq:intIL}
\end{eqnarray} 
Using the $\boldsymbol p_{2 T}$ integration to eliminate the delta function 
in Eq.\ (\ref{eq:intIL}) 
and shifting the integration variable $\boldsymbol p_{1 T} 
\to\boldsymbol p_{1 T}' = \boldsymbol p_{1 T} -\frac{1}{2}\boldsymbol q_T$,
one arrives at
\begin{eqnarray}
 \int d^2 \boldsymbol q_T \,\boldsymbol{q}_T^4\, (2 {\cal I}^{g g} -{\cal L}^{g g }) & = &
\frac{1}{M_1^2 M_2^2}\frac{R^4}{16\, \pi} \int d^2 \boldsymbol q_T \,
d \boldsymbol p_{1 T}'^2\, \boldsymbol q_T^4\, e^{- R^2 
\big (2 \boldsymbol p_{1 T}'^2 + \frac{1}{2} \boldsymbol q_T^2 \big )}
\,h_1^{\perp g}(x_1) h_1^{\perp g}(x_2) \nonumber \\
&=& \frac{1}{2 M_1^2 M_2^2} \,\frac{1}{R^4}\, 
h_1^{\perp g}(x_1) h_1^{\perp g}(x_2)~.
\label{eq:integral} 
\end{eqnarray}
Substituting Eq.\ (\ref{eq:integral}) into Eq.\ (\ref{eq:weight}) 
shows that for a Gaussian shape one finds for
the unweighted average
the final result
\begin{eqnarray}
\pi \int d\boldsymbol{q}_T^2\ \langle \cos 4 (\phi_T-\phi_\perp)  \rangle 
& = & \frac{\alpha_s^2}{s  \boldsymbol K_{\perp} ^2 } \, \frac{N}{N^2-1}
\, {y (1-y)} \bigg [ N (1-y (1-y)) -\frac{1}{4} 
\bigg (y^2 + (1-y)^2-\frac{1}{N^2} \bigg )\bigg ] 
\nonumber \\ && 
\qquad\qquad\qquad \mbox{} \times
h_1^{\perp g (1)}(x_1) 
h_1^{\perp g (1)}(x_2). 
\end{eqnarray}

%%%%%%%%%%%%%%%%%%%%%%%%%%%%%%%%%%%%%%%%%%%%%%%%%%%%%%%%%%%%%%%%%%%%%%%%%%%%
\section{\label{broadening}
Jet broadening}
%%%%%%%%%%%%%%%%%%%%%%%%%%%%%%%%%%%%%%%%%%%%%%%%%%%%%%%%%%%%%%%%%%%%%%%%%%%%

In our almost back-to-back jet situation, the jet-direction $j$
coincides with the transverse thrust axis and the
jet broadening variable $Q_t$ defined in Eq.\ (\ref{eq:Qt}) is given by
\begin{equation} 
Q_t =  E_T\,\left |\sin \left ( \frac{\delta\phi}{2}\right )\right | 
= |\boldsymbol{q}_T||\sin(\phi_T-\phi_j)|
\approx
|\boldsymbol{q}_T||\sin(\phi_T-\phi_\perp)| \, ,
\label{eq:Qt2}
\end{equation}
for which we refer to Eq.\ (\ref{two}) in Appendix~\ref{appendixA} 
and one needs to use Eqs.\ (\ref{three}) and (\ref{four}) to check the
validity of the approximation. Using this expression we can
now turn  to the evaluation of the average jet broadening 
$\langle Q_t \rangle$ as a function of $|\boldsymbol K_{\perp}|$,
\begin{equation}
\langle Q_t \rangle  \propto  \int \frac{d\phi_\perp}{2\pi} 
\,d^2 \boldsymbol{q}_T 
\ Q_t \left (|\boldsymbol q_T|, \phi_T, \phi_\perp\right )\, 
\frac{d\sigma^{h_1 h_2 \rightarrow {\rm jet} \,{\rm jet} \, X}}
{d^2\boldsymbol{K}_{\perp}\,d^2\boldsymbol{q}_{T}}~.
\end{equation}
The differential cross section in 
the integrand is obtained from Eq.\ (\ref{eq:cso}) and contains, besides the 
well-known, angular independent term $A$, also the terms $B$ (due to
the transverse polarization of quarks and antiquarks in the colliding hadrons)
and $C$ (related to the linear polarization of gluons).   
The following integrals,
\begin{eqnarray}
&&
\int_0^{2 \pi} d\phi_T\, |\sin (\phi_T-\phi_\perp)|  = 4 \, ,
\\ &&
\int_0^{2 \pi} d\phi_T\, |\sin (\phi_T-\phi_\perp)| 
\cos 2 (\phi_T-\phi_\perp) = -\frac{4}{3} \, ,
\\ &&
\int_0^{2 \pi}  d\phi_T\,|\sin (\phi_T-\phi_\perp)| 
\cos 4 (\phi_T-\phi_\perp)  = -\frac{4}{15}\, 
\end{eqnarray}
are all different from zero, meaning that the $A$, $B$ and $C$ terms 
contribute to 
$\langle Q_t\rangle$: 
\begin{equation}
\langle Q_t \rangle  \propto  \int d^2 \boldsymbol{q}_T |\boldsymbol q_T|
\bigg [ A(\boldsymbol{q}_T^2) 
- \frac{1}{3} {B}(\boldsymbol{q}_T^2) \boldsymbol q_T^2 -
\frac{1}{15} {C}(\boldsymbol{q}_T^2) \boldsymbol{q}_T^4 \bigg ]~.
\end{equation}

In order to calculate $\langle Q_t^2\rangle$, one
needs to evaluate the integrals  
\begin{eqnarray}
&&
\int_0^{2 \pi} d\phi_T\, \sin^2 (\phi_T-\phi_\perp) = \pi  \, ,
\\ &&
\int_0^{2 \pi} d\phi_T\, \sin^2 (\phi_T-\phi_\perp) 
\cos 2 (\phi_T-\phi_\perp)  = -\frac{\pi}{2}  \, ,
\\ &&
\int_0^{2 \pi} d\phi_T\, \sin^2 (\phi_T-\phi_\perp) 
\cos 4 (\phi_T-\phi_\perp)  = 0\,, 
\end{eqnarray}
which show that only the terms $A$ and $B$ enter in the estimate of 
$\langle Q_t^2\rangle$: 
\begin{equation}
\langle Q_t^2 \rangle  \propto  \int d^2 \boldsymbol{q}_T \boldsymbol q_T^2
\bigg [ A(\boldsymbol{q}_T^2) 
- \frac{1}{2} {B}(\boldsymbol{q}_T^2) \boldsymbol q_T^2 \bigg ]~.
\end{equation}

%%%%%%%%%%%%%%%%%%%%%%%%%%%%%%%%%%%%%%%%%%%%%%%%%%%%%%%%%%%%%%%%%%%%%%%%%%%%
%%%%%%%%%%%%%%%%%%%%%%%%%%%%%%%%%%%%%%%%%%%%%%%%%%%%%%%%%%%%%%%%%%%%%%%%%%%%

\section{\label{section6}
Color flow dependence and factorization}
 
In our treatment in this paper we have simply convoluted the quark and gluon
correlators with the hard partonic cross sections, without worrying
about possible nontrivial effects arising from the gauge link
structure in these correlators. The proper gauge invariant
definitions of TMDs as well as collinear correlators
involve nonlocal operators containing path-ordered exponentials, 
the gauge links. The gauge link is the result of resumming all
gluons with polarizations along the momentum of a particular hadron
into the soft parts. In the case of TMDs the path
of the gauge links generally depends on the process.
The path dependence disappears after integration over transverse momenta.
In the collinear correlators, one can usually choose a
gauge that makes the gauge link unity, but the same procedure
for TMDs can leave transverse pieces that are situated at lightcone
infinity. These links can have physical effects, for instance 
in single transverse spin asymmetries that arise from the Sivers effect,
which is described by a T-odd TMD. The Sivers asymmetries in
semi-inclusive deep inelastic scattering and the Drell-Yan process 
are predicted to differ
by a sign as a consequence of the gauge links \cite{Collins:2002kn}.

In the more complicated processes $h_1^\uparrow h_2 \rightarrow
\gamma/{\rm jet} + {\rm jet} + X$ the single spin asymmetries involving 
the Sivers function~\cite{Bacchetta:2007sz,Boer:2003tx,Bomhof:2007su} come
from correlators with more complex paths in the gauge links. This 
causes deviations that are more involved than a simple sign change with 
respect to e.g.\ semi-inclusive deep inelastic scattering. 
But also in this case, calculable process-dependent ``color flow'' factors
can be obtained which may be different for each hard partonic subprocess.
In this way they allow for the calculation of particular
weighted cross sections in dijet production, resulting in a small asymmetry
\cite{Bomhof:2007su,Qiu:2007ey,Vogelsang:2007jk}, as also shown
by the data \cite{Abelev:2007ii}. However, claims of possible
factorization breaking have been put forward for this process
\cite{Collins:2007nk,Collins:2007jp} and this remains an open question.
  
For observables involving a product of two T-odd TMDs, such as the
one discussed in the present paper, the situation is less
clear. For $\cos 2\phi$ asymmetries in Drell-Yan \cite{Boer:1999mm} 
and $h_1 h_2 \to \gamma \, {\rm jet}\, X$ 
the effects of nontrivial gauge links were included in Ref.\  
\cite{Boer:2007nd} following the methods outlined in 
Refs.\ \cite{Bomhof:2006dp,Pijlman:2006tq,Bomhof:2006ra,Bomhof:2007xt}. 
In both cases the color flow factor obtained was $+1$.
However, since
the methods used were developed for observables involving a single
non-contracted transverse momentum $p_T$ for a T-odd TMD for one of the
hadrons in the process, the extension to cases in which non-contracted
transverse momenta of partons in two different hadrons are involved 
certainly needs careful study. 
The $p_T$-dependence for $h_1^\perp$ in the correlators 
for gluons, moreover, has a rank two tensor structure in the non-contracted
transverse momentum, although it is T-even.
For the present case of dijet production (for which
nontrivial color flow factors were presented in Ref.\ \cite{Lu:2008qu}),
which is necessarily more complicated and for which doubts about
factorization have been put forward, at this stage we do not include
any color flow factors. Since we have presented the expressions
for each partonic subprocess separately, it is possible to include
the correct factors at a later stage, once they have been
firmly established. If factorization cannot be proven for the process of
interest, however, this not only implies that the functions $h_1^\perp$ cannot 
be extracted but neither can in that case $\langle p_T^2 \rangle$ be obtained.

%%%%%%%%%%%%%%%%%%%%%%%%%%%%%%%%%%%%%%%%%%%%%%%%%%%%%%%%%%%%%%%%%%%%%%%%%%%%%
%%%%%%%%%%%%%%%%%%%%%%%%%%%%%%%%%%%%%%%%%%%%%%%%%%%%%%%%%%%%%%%%%%%%%%%%%%%%%

\section{Summary and Conclusions}

In this paper we study the effects of transverse momenta of the
initial state hadrons in hadronic dijet production. The transverse
momentum produces an imbalance in the dijets in the transverse plane.
In the usual treatments the effects are attributed to gluon radiation
and to the transverse momentum dependence of the unpolarized quark
distributions.  We look at the effects of two additional TMD functions
that enter in the scattering of unpolarized hadrons, the distribution
of transversely polarized quarks ($h_1^{\perp q}$) and the
distribution of linearly polarized gluons ($h_1^{\perp g}$). They
produce specific azimuthal dependences of the two jets, that are
  not suppressed {\it a priori}. The effects
cannot be isolated by only looking at the angular deviation from
the back-to-back situation, but depend on the jet transverse energy
and the contributions to it of the two jets. We have discussed
appropriate weighting to isolate the specific additional
contributions. We also pointed out their effect on the jet
  broadening quantities $\langle Q_t \rangle$ and $\langle
  Q_t^2\rangle$, which we considered for the simplest two-jet case,
  but the conclusion that $h_1^\perp$ functions contribute to them 
  also affects the more general cases containing a sum over hadrons. This
 in principle complicates the extraction of the
average initial parton transverse momentum from the jet broadening,
but possibly even hampers it altogether if factorization of the
(sufficiently sizeable) spin
dependent contributions indeed turns out to be broken.

%%%%%%%%%%%%%%%%%%%%%%%%%%%%%%%%%%%%%%%%%%%%%%%%%%%%%%%%%%%%%%%%%%%%%%%%%%%%%
%%%%%%%%%%%%%%%%%%%%%%%%%%%%%%%%%%%%%%%%%%%%%%%%%%%%%%%%%%%%%%%%%%%%%%%%%%%%%
\section*{ACKNOWLEDGMENTS}

We thank Ted Rogers for useful discussions. 
This research is part of the research program of the 
``Stichting voor Fundamenteel Onderzoek der Materie (FOM)'', 
which is financially supported by the ``Nederlandse Organisatie voor Wetenschappelijk Onderzoek (NWO)''. 

%%%%%%%%%%%%%%%%%%%%%%%%%%%%%%%%%%%%%%%%%%%%%%%%%%%%%%%%%%%%%%%%%%%%%%%%%%%%%
%%%%%%%%%%%%%%%%%%%%%%%%%%%%%%%%%%%%%%%%%%%%%%%%%%%%%%%%%%%%%%%%%%%%%%%%%%%%%
\appendix
\section{\label{appendixA}
Transverse plane variables}

In the transverse plane we have the two jet momenta $\boldsymbol{K}_{1\perp}$  
and $\boldsymbol{K}_{2\perp}$ defining azimuthal angles $\phi_1$ and $\phi_2$.
\mbox{From} them one can construct the sum and difference angles,
\begin{eqnarray}
&&
\phi_j = (\phi_1 + \phi_2 -\pi)/2,
\\&&
\delta\phi = \phi_1 - \phi_2 -\pi.
\end{eqnarray}
The sum and difference of the transverse energies of the two jets, 
$\vert \boldsymbol{K}_{1\perp} \vert$ and $\vert
\boldsymbol{K}_{1\perp}\vert$, define
\begin{eqnarray}
&&
E_T 
= \vert\boldsymbol{K}_{1\perp}\vert + \vert\boldsymbol{K}_{2\perp}\vert
\\&&
\Delta K_\perp 
= \vert\boldsymbol{K}_{1\perp}\vert - \vert\boldsymbol{K}_{2\perp}\vert
\end{eqnarray}
\begin{figure}[h]
\epsfig{figure=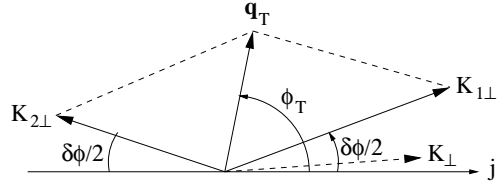,width=6.5cm} 
\caption{\label{transverseplane}
The transverse plane is defined as orthogonal with respect to the two
incoming hadrons. The jet direction ($j$) is defined as $\phi_j =
(\phi_1 + \phi_2 - \pi)/2$. The momenta 
$\boldsymbol{q}_T = \boldsymbol{K}_{1\perp} + \boldsymbol{K}_{2\perp}$ and 
$\boldsymbol{K}_\perp =(\boldsymbol{K}_{1\perp} - \boldsymbol{K}_{2\perp})/2$
define the azimuthal angles $\phi_T$ and $\phi_\perp$.}
\end{figure}
One can use
$d^2\boldsymbol{K}_{1\perp}\,d^2\boldsymbol{K}_{2\perp}$ =
$\tfrac{1}{8}\,(E_T^2-\Delta K_\perp^2)\,dE_T\,d\Delta K_\perp
\,d\phi_j\,d\delta\phi$ 
for the phase space or go to the sum and difference momenta and their angles
as shown in Fig.~\ref{transverseplane}. In that case one has
$d^2\boldsymbol{K}_{1\perp}\,d^2\boldsymbol{K}_{2\perp}$ =
$d^2\boldsymbol{K}_{\perp}\,d^2\boldsymbol{q}_{T}$ =
$\vert\boldsymbol{q}_T\vert \,\vert\boldsymbol{K}_\perp\vert
d\vert\boldsymbol{q}_T\vert\,d\vert\boldsymbol{K}_\perp\vert
\,d\phi_T\,d\phi_\perp$.
We have the following exact relations
\begin{eqnarray}
&& 
\vert\boldsymbol{q}_T\vert \,\cos(\phi_T-\phi_j) 
= \Delta K_\perp\,\cos (\delta\phi/2),
\\&&
\vert\boldsymbol{q}_T\vert \,\sin(\phi_T-\phi_j) 
= E_T\,\sin (\delta\phi/2),
\label{two}
\\&&
2\,\vert\boldsymbol{K}_\perp\vert \,\cos(\phi_\perp-\phi_j) 
= E_T\,\cos (\delta\phi/2),
\label{three}
\\&&
2\,\vert\boldsymbol{K}_\perp\vert \,\sin(\phi_\perp-\phi_j) 
= \Delta K_\perp\,\sin (\delta\phi/2),
\label{four}
\\&&
\boldsymbol{q}_T^2 =
\Delta K_\perp^2\,\cos^2 (\delta\phi/2)
+ E_T^2\,\sin^2 (\delta\phi/2)
= \Delta K_\perp^2
+ (E_T^2-\Delta K_\perp^2)\,\sin^2 (\delta\phi/2),
\label{five}
\\&&
4\,\boldsymbol{K}_\perp^2
=E_T^2\,\cos^2 (\delta\phi/2)
+ \Delta K_\perp^2\,\sin^2 (\delta\phi/2)
=E_T^2
- (E_T^2 - \Delta K_\perp^2)\,\sin^2 (\delta\phi/2),
\label{six}
\\&&
2 \vert\boldsymbol{q}_T\vert \,\vert\boldsymbol{K}_\perp\vert 
\,\cos(\phi_T-\phi_\perp) = E_T\,\Delta K_\perp, 
\label{seven}
\\&&
2 \vert\boldsymbol{q}_T\vert \,\vert\boldsymbol{K}_\perp\vert 
\,\sin(\phi_T-\phi_\perp) = (E_T^2-\Delta K_\perp^2)/2\,\sin(\delta\phi).
\label{eight}
\end{eqnarray}
We note that the order of the momenta is 
$\vert\boldsymbol{q}_T\vert \sim 
\Delta K_\perp \sim M$ (hadronic scale) while
$E_T \sim 2\,\vert\boldsymbol{K}_\perp\vert \sim \sqrt{s}$, so we see
from Eq.\ (\ref{two}) that $\delta\phi \sim M/\sqrt{s}$ and
from Eq.\ (\ref{four}) that $\phi_\perp-\phi_j \sim M^2/s$.
Further useful relations are 
\begin{eqnarray}
&&
\vert\boldsymbol{q}_T\vert^2\,\cos 2(\phi_T-\phi_\perp)
\approx \vert\boldsymbol{q}_T\vert^2\,\cos 2(\phi_T-\phi_j)
= \Delta K_\perp^2\,\cos^2(\delta\phi/2) - E_T^2\,\sin^2(\delta\phi/2),
\\&&
\vert\boldsymbol{q}_T\vert^2\,\sin 2(\phi_T-\phi_\perp)
\approx \vert\boldsymbol{q}_T\vert^2\,\sin 2(\phi_T-\phi_j)
= E_T\,\Delta K_\perp\,\sin(\delta\phi),
\\&&
\vert\boldsymbol{q}_T\vert^4\,\cos 4(\phi_T-\phi_\perp)
\approx \vert\boldsymbol{q}_T\vert^4\,\cos 4(\phi_T-\phi_j)
\approx \Delta K_\perp^4 + E_T^4\,\sin^4(\delta\phi/2)
- 6\,E_T^2\,\Delta K_\perp^2\,\sin^2(\delta\phi/2).
\end{eqnarray}

\section{\label{appendixB}
Photon-jet production}

 In this appendix we include expressions for the terms 
  $A$ and $B$ for the photon-jet production case, because in Ref.\
  \cite{Boer:2007nd} we only considered approximate angular dependence. 
Similarly to Eq.\ (\ref{eq:cso}), one can write 
\begin{equation}
\frac{d\sigma^{h_1 h_2 \rightarrow {\gamma} \,{\rm jet} \, X}}{d\eta_\gamma d\eta_j d^2\boldsymbol{K}_{\gamma\perp} d^2\boldsymbol{K}_{j \perp}} =
\frac{\alpha \alpha_s}{s \boldsymbol{K}_\perp^2}
\bigg[ A(\boldsymbol{q}_T^2) + B(\boldsymbol{q}_T^2) \boldsymbol q_T^2 \cos 2 (\phi_T-\phi_\perp) \bigg ]\, ,
\label{eq:csogamma}
\end{equation}
with 
\begin{equation}
A(y, x_1, x_2, \boldsymbol q_T^2) = {\cal A}^{q g\to \gamma q}  + {\cal A}^{q \bar q \to \gamma g } \,,\qquad  
B(y, x_1, x_2, \boldsymbol q_T^2) = {\cal B}^{q \bar q \to \gamma g }~.
\end{equation}
By comparison with Eqs.\ (15), (16), and (19) in Ref.\ \cite{Boer:2007nd}, we find the following 
expressions
\begin{eqnarray}
{\cal A}^{q g\to \gamma q} = \sum_{q} e^2_q \bigg [ h(y) {\cal F}^{q g}(x_1, x_2,\boldsymbol{q}_T^2) 
+ h(1-y) {\cal F}^{g q}(x_1, x_2,\boldsymbol{q}_T^2) \bigg ]\, ,
\end{eqnarray}
with
\begin{equation}
h(y) = \frac{1}{N} (1-y)(1+y^2)\, ,
\end{equation}
\begin{eqnarray}
{\cal A}^{q \bar q \to \gamma g }&  = & \frac{N^2-1}{N^2} \, (y^2 + (1-y)^2) \sum_{q} e^2_q \bigg [
{\cal F}^{q \bar q}(x_1, x_2,\boldsymbol{q}_T^2) + {\cal F}^{\bar q  q}(x_1, x_2,\boldsymbol{q}_T^2)\bigg ]\, ,
\end{eqnarray}
and 
\begin{equation}
{\cal B}^{q \bar q\to \gamma g } = 2\, \frac{N^2-1}{N^2}\, y (1-y) \sum_{q} e^2_q 
\bigg [{\cal H}^{q \bar q}(x_1, x_2,\boldsymbol{q}_T^2) + {\cal H}^{\bar q  q}(x_1, x_2,\boldsymbol{q}_T^2) \bigg ]\, ,
\end{equation}
in agreement with the results in Refs.\ \cite{Ji:1992ev,Jaffe:1996ik,Soffer:2002tf}.
%%%%%%%%%%%%%%%%%%%%%%%%%%%%%%%%%%%%%%%%%%%%%%%%%%%%%%%%%%%%%%%%%%%%%%%%%%%%%
%%%%%%%%%%%%%%%%%%%%%%%%%%%%%%%%%%%%%%%%%%%%%%%%%%%%%%%%%%%%%%%%%%%%%%%%%%%%%

\end{document}